\documentclass[sn-mathphys,Numbered]{sn-jnl}

\usepackage{graphicx}%
\usepackage{multirow}%
\usepackage{amsmath,amssymb,amsfonts}%
\usepackage{amsthm}%
\usepackage{mathrsfs}%
\usepackage[title]{appendix}%
\usepackage{xcolor}%
\usepackage{textcomp}%
\usepackage{manyfoot}%
\usepackage{booktabs}%
\usepackage{algorithm}%
\usepackage{algorithmicx}%
\usepackage{algpseudocode}%
\usepackage{listings}%

\theoremstyle{thmstyleone}%
\theoremstyle{thmstyletwo}%

\theoremstyle{thmstylethree}%

\begin{document}

\title[I-BaR: Integrated Balance Rehabilitation Framework]{I-BaR: Integrated Balance Rehabilitation Framework}


\author*[1]{\fnm{Tugce} \sur{Ersoy}}\email{tugce.ersoy@ozu.edu.tr}

\author[2]{\fnm{Pınar} \sur{Kaya}}\email{pkaya@medipol.edu.tr}
\equalcont{These authors contributed equally to this work.}

\author[3,4]{\fnm{Elif} \sur{Hocaoglu}}\email{ehocaoglu@medipol.edu.tr}
\equalcont{These authors contributed equally to this work.}
\author[1]{\fnm{Ramazan} \sur{Unal}}\email{ramazan.unal@ozyegin.edu.tr}
\equalcont{These authors contributed equally to this work.}

\affil*[1]{\orgdiv{Mechanical Engineering}, \orgname{Ozyegin University}, \orgaddress{\street{Orman}, \city{Istanbul}, \postcode{34794},  \country{Türkiye}}}

\affil[2]{\orgdiv{Physiotherapy and Rehabilitation}, \orgname{Istanbul Medipol University}, \orgaddress{\street{Ekinciler}, \city{Istanbul}, \postcode{34810},  \country{Türkiye}}}

\affil[3]{\orgdiv{Electrical and Electronics Engineering}, \orgname{Istanbul Medipol University}, \orgaddress{\street{Ekinciler}, \city{Istanbul}, \postcode{34810}, \country{Türkiye}}}

\affil[4]{\orgdiv{SABITA (Research Institute for Health Sciences and Technologies)}, \orgname{Istanbul Medipol University}, \orgaddress{\street{Ekinciler}, \city{Istanbul}, \postcode{34810}, \country{Türkiye}}}


\abstract{
\textbf{{Background \& Methods}} 
Neurological diseases are observed in approximately one billion people worldwide. A further increase is foreseen at the global level as a result of population growth and aging. Individuals with neurological disorders often experience cognitive, motor, sensory, and lower extremity dysfunctions. Thus, the possibility of falling and balance problems arise due to the postural control deficiencies that occur as a result of the deterioration in the integration of multi-sensory information.

\textbf{{Result}} 
we propose a novel rehabilitation framework, Integrated Balance Rehabilitation (I-BaR), to improve the effectiveness of the rehabilitation with objective assessment, individualized therapy, convenience with different disability levels and adoption of assist-as-needed paradigm and, with integrated rehabilitation process as whole, i.e., ankle-foot preparation, balance, and stepping phases, respectively. 

\textbf{{Conclusions}} 
Integrated Balance Rehabilitation allows patients to improve their balance ability by providing multi-modal feedback: visual via utilization of Virtual Reality; vestibular via anteroposterior and mediolateral perturbations with the robotic platform; proprioceptive via haptic feedback.
}

\maketitle

\section{Background}
Neurological disorders impact approximately 1 billion individuals worldwide, representing a diverse range of socio-economic statuses, age groups, and ethnicities. Furthermore, an estimated 6.8 million people die each year as a result of neurological illnesses \cite{Who_2022, Fineberg_2013, Pehlivan_2016}. The number of deaths from neurological disorders and disability has increased dramatically in the last 30 years, particularly in low- and middle-income countries; a further increase is foreseen worldwide due to growth in population and aging. Neurological disorders are predicted to impacting for about  10\% of the global burden of disease and 25\% of years lived with disability \cite{Who_2022}. Along with the cost of care for neurological disorders, the financial loss related to unemployment etc., was estimated to be \$9 billion per year in Canada in 2007 \cite{Gaskin_2017}, and about €1135 million in the United Kingdom in 2010 \cite{Fineberg_2013}. In the United States, it was estimated that expenses only for stroke alone exceeded \$33.6 billion in 2015 \cite{Pehlivan_2016}. In order to provide adequate service in this regard, it is vital to take action to meet the growing demand.  

Individuals with neurological disorders experience walking and movement restrictions, balance, motor control, exhaustion, and other health problems, which all directly impact their quality of life (QoL) due to impairments in the central and peripheral nervous systems (CNS and PNS). The nature and severity of the condition may differ from patient to patient depending on the lesional region at the affected CNS and PNS. In Multiple Sclerosis (MS), commands are conveyed through the nerve slowly or cannot be transmitted due to degeneration of the myelin sheath \cite{Mehravar_2015,Doty_2018}. In Parkinson's disease (PD), depigmentation of the substantia nigra and damage to dopamine-producing cells result in deficiencies in the balance control \cite{Deng_2018}. Further, in stroke disease, focal neurological function losses occur due to infarction or bleeding in the relevant part of the brain \cite{Hankey_2017}. The CNS integrates information from the visual, vestibular, proprioceptive and cognitive systems, and PNS through continuous sensory re-weighting; provides postural control in static and dynamic situations \cite{Horak_2006}. This integration of multi-sensory information is disrupted in neurological diseases, causing balance deficits and an increase in fall rate \cite{Rito_2021}. More than 75\% of MS patients have symptoms of poor balance. Further, approximately 60\% of MS patients reported at least one fall in the previous three months, and more than 80\% reported impairments in activities of daily living (ADL) \cite{Aruin_2015a,Craig_2019}. It has been reported that approximately 73\% of stroke patients and 45-68\% of Parkinson's patients experience at least one fall per year \cite{De_2021}. The factors mentioned above lead to a sedentary lifestyle and accordingly deteriorate patients' social health, which is a subdomain of QoL. This lifestyle may also lead to additional health-related issues such as obesity, diabetes, and heart disease \cite{Craig_2019, Schilling_2019}, which further dramatically deteriorates their QoL \cite{PMI}. Due to the factors mentioned above, the search for improvements in existing rehabilitation methods continues.

 With this aim, we identify the necessity of integrated balance rehabilitation (I-BaR) framework to assess and improve ankle-foot proprioception, postural control, and stepping characteristics in neurological diseases. In particular, this framework is composed of three main phases, i.e., ankle-foot/preparation, balance and stepping rehabilitation. At first, in the ankle-foot/preparation phase, the sensation of the sole, joint proprioception, and movement improvements are aimed. Secondly, in the balance phase, sensory weighting skills are aimed to be improved by using multi-modal feedback via perturbations. Lastly, stepping rehabilitation aims to improve walking parameters via step-taking activities to target points with adjustable distances.

\section{Methods}
The skill of maintaining balance involves multiple factors that combine both physical and sensory elements. The current balance rehabilitation program is used different sequences of training, feedback, assistance, instruction, and focus of attention, as well as exercise physiology principles. These devices/tests have been developed for different disease types and severity levels. However, since these devices/tests address specific severity levels, they cannot be used in the entire rehabilitation process. To the best of our knowledge, there is no methodology developed for ankle, balance, and step-taking rehabilitation for improving postural adjustment strategies (motor learning) in patients with different disease severity. The following section presents the current finding and solution of the overview of postural control mechanisms, robot-aided rehabilitation, and their design, respectively.

\subsection{Overview of postural control mechanisms}
The somatosensory function includes the senses of touch, vibration, pressure, proprioception, pain, and temperature. Impairments in this function negatively affect the ability to perceive, distinguish and recognize the senses in the body \cite{Aries_2021}. Consequently, disorientation accompanied by abnormal movements, balance disorders, muscle weakness, and inability to maintain postural control (stabilize the body against gravity and perturbation) may occur \cite{Kim_2021}. For instance, it is reported that post-stroke individuals experience high rates of somatosensory impairment, ranging between 65-85\% \cite{Costantino_2017}. A research with self-questionnaire demonstrated that individuals with somatosensory and motor impairments suffer from lower walking capacities and lower levels of independence in ADL \cite{Gorst_2019, Tyson_2013, Patel_2000}. Since restoring walking ability is a primary objective for many stroke patients, establishing the best treatments for balance, gait, and mobility were identified as one of the top ten stroke research priorities \cite{hez_1999}. In a recent survey conducted with 145 stroke individuals, 43\% of individuals reported decreased sensation in their feet; sensory impairment was indicated to be the second most common foot problem after the loss of strength \cite{Gorst_2016, Bowen_2016}. Limitations in walking, high fall rate, and impairments in foot-ground contact and sense of foot position sense and hence, decrease the outdoor activities in the community.

\subsubsection{Balance control mechanisms}
Balance control, according to Shumway-Cook and Woollacott \cite{Anne_2016}, is highly activity-specific and falls into three categories:
\begin{itemize}
    \item \emph{{Static/dynamic steady-state balance control:}} is sustaining a stable posture while sitting, standing, or walking.
    \item \emph{{Proactive balance control}}: is activated before the predicted perturbation. The CNS uses postural adjustments, i.e., anticipatory and compensatory postural adjustments (APAs/CPAs), envisaged as a muscular adjustment mechanism to provide balance control while maintaining body balance and vertical posture during different conditions. These adjustments engage and activate the trunk and lower extremity postural muscles before an impending external/internal perturbation occurs. It reduces the risk of balance deterioration by regulating the body's center of mass (CoM) position.
    \item \emph{{Reactive balance control}}: is activated after the perturbation to compensate CoM deviation \cite{Granacher_2011,Lesinski_2015}. The CNS uses CPAs as a muscular adjustment mechanism to provide this balance control. These adjustments are triggered by the sensory control signal and allow the CoM to be repositioned once it is disturbed \cite{Aruin_2015b}.
\end{itemize}

After people lose their balance, they only have a few seconds to coordinate and stabilize their posture \cite{Horak_2006, Aruin_2017b}. Postural perturbations such as sliding and tripping in everyday situations vary widely and are highly unpredictable. Recently, it has been proposed that studying the processes of APAs and CPAs may reveal vital information about postural control and falls. Numerous research findings show that due to postural deficiencies, falls occur during ADL \cite{Tajali_2018, Krishnan_2012, Shadmehr_2012}. The duration and magnitude of muscular activation were measured in different studies, and significant APAs deficiencies were identified. There is evidence that improvements in the production of APAs can be achieved even after a single training session in individuals with stroke \cite{Aruin_2017b}. In another study \cite{Aruin_2016} with elderly healthy individuals shows that APAs improvement can also be achieved after four weeks of external perturbation training. These studies show that individuals exposed to predicted perturbations provide better compensatory activity in the muscles (improvement in APAs and CPAs) and more adequate body pressure center changes with the use and production of strong APAs \cite{Aruin_2017b, Aruin_2016}. After training, early muscles activation and reduced CoM excursions took place which is substantial evidence that retraining of APAs is possible. These results form an essential basis for investigating training effectiveness focused on improving long-term APAs, CPAs, and reaction time in increasing individuals' postural control. Furthermore, they provide a background for the development of perturbation programs to improve postural control, balance and prevent falls.

Reaction time is commonly considered in clinical diagnosis, treatment, and follow-up stages to determine the severity of postural control deficiency (APAs and CPAs) in somatosensory-based motor and neurological disorders \cite{Saito_2014, Sandroff_2015}. It is the time elapsed between the onset of a stimulus and when the patient's response to that stimulus begins which is physiologically divided into five parts. These are; (1) seeing the stimulus at the receptor level, (2) transmitting the stimulus to the CNS, (3) transmitting the stimulus through the nerves, (4) generating the effector signal, (5) transporting the signal to the muscles through the CNS for the mechanical work to be done \cite{AGIRBAS_2019, Tajali_2019, Dejanovic_2015}. When they examined the surface electromyography (sEMG) activity of the lower extremity muscles and the center of pressure (CoP) of both falling and non-falling stroke patients, it was found that the falling group had lower muscular electrical activity. Therefore, they need longer reaction times to prepare the posture and initiate movements through APAs \cite{Santos_2010}.

The reactive and proactive balance control are managed by the activation of different neurological mechanisms of the CNS. Impairment in one of the postural adjustments (APAs and CPAs), and their ability to affect each other negatively that highlights the importance of assessing and training them in rehabilitation. Loss of balance may occur due to an unpredictable external force or failure of balance control after external/internal perturbation, i.e., fast and voluntary extremities movement. Therefore, it is possible to observe improvements in both postural adjustments (proactive and reactive balance control) with effective fall prevention training \cite{Aruin_2015, Yamada_2021, Aruin_2017a}.

A novel approach to analyze APAs employs Fitts' law to explain the relationship between APAs parameters, the length of the step, and the size of the stepping target. It is a valuable method for various target-directed movements and quantifying stability control factors. Thus, previous literature on able-bodied individuals report that Fitts' law is a valid way to explain the time to complete the foot-reaching task and APAs levels \cite{Aloraini_2020, Bertucco_2013, Bertucco_2010, Mulder_2013}. A voluntary step initiation is a self-perturbation of balance with a change in the base of support and the transition from a static to a dynamic posture, so the velocity and accuracy of movement can be assessed with Fitts’ tasks since coordinated muscle activation prior to voluntary movement (APAs) are utilized to maintain the posture. A Choice Stepping Reaction Time (CSRT) test is a simple activity that evaluates person's ability to immediately trigger and execute a step with Fitts' law. The subject must step on one of the numerous targets put in front of or around them as rapidly as feasible. The time it takes to attain the goals is a promising strategy for assessing fall risk among the elderly population since they have a significantly longer duration in reaction time than non-fallers. Furthermore, a few studies investigated the inverse proportion between speed and accuracy control in foot-reaching tasks. Patients were asked to use their affected leg to step to targets with different sizes and at varying distances during these tasks \cite{Tajali_2019, Yamada_2021, Barr_2014}. However, these target points only include a switch button to calculate the reaction time. In other words, they cannot measure ground reaction force (GRF) and give haptic feedback to the patient. Yet, as neurological patients report decreased sensation on the sole and trouble in weight shifting, so it is vital to measure GRF for appropriate feedback and address aforementioned challenges in the training \cite{Chien_2017}. 

\subsubsection{The effects of perturbation-based balance training} \label{sec:The effects of perturbation-based balancing training}
Perturbation-based balance training (PBT) is a type of exercise in which participants are intentionally disturbed to improve reactive balance reactions by training the individual neuromuscular responses \cite{Mansfield_2015, Allin_2020, Barzideh_2020}. This training requires performing rapidly occurring sequential whole-body movements and applying large and sudden disruptive forces to stabilize CoM \cite{Pai_2014}. With the development of balance reactions, an increase in the ability to respond to the loss of balance in unpredictable ADL and consequently a decrease in fall rate can be achieved \cite{Mansfield_2015}. 

In addition to the physiotherapist's manual pushes and pulls (lean and release test) in PBT studies, treadmill acceleration-deceleration and inclined/moving platforms have been implemented in recent years to mimic external perturbations in daily life \cite{Pai_2014, Bhatt_Pai_2008, Bhatt_Pai_2009}. It is suggested that perturbation training while walking could be an effective way to minimize fall rates in elderly people \cite{Gerards_2017, Pai_2007}. Current research suggests that CPAs in the elderly can be improved using PBT and that these improvements can be sustained for up to a full year after training \cite{Gerards_2017}. In another study, it is shown that only a single session of perturbation is sufficient to provide permanent improvements in reactive balance control and prevent falls in elderly individuals \cite{Aruin_2017b}. However, to the best of our knowledge, no study has been conducted on the optimal dosage of perturbation training to induce permanent changes in reactive balance control.

Although the PBT is an approach to decrease the fall rate, it is still far from the realistic condition. The limited type of perturbations performed with existing devices and techniques may reduce individuals' capacity of adapting and generalizing the effects of PBT training to ADL. On top of that different perturbation modalities in PBT programs can be considered highly important to train balance reactions to match with a variety of situations and motor tasks \cite{Mansfield_2011, Aviles_2020, Tanvi_2012}.

\subsubsection{Proprioceptive/sensorimotor training}\label{sec:Stroke proprioceptive/sensorimotor training}
Sensory-motor training gradually improves an individual’s ability to re-weight and integrate sensory inputs in order to control balance and prevent falls in different somatosensory input situations \cite{Gandolfi_2015}. According to these approaches, new technological devices and paradigms are being developed to promote neurorehabilitation from the CNS to the PNS. Moreover, the participation of cognitive functions increased by integrating multi-sensory feedback thus it improve rehabilitation effectiveness \cite{Morone_2019, Verna_2020, Kearney_2019}.

Smania et al. \cite{Smania_2008} show significant improvements in the ability of stroke patients to maintain balance control with a unique training program based on weight transfer and balance exercises performed under different manipulation of sensory inputs. Derakhshanfar et al. \cite{Derakhshanfar_2021} report that exteroceptive and proprioceptive stimulations, which include sensory intervention, are effective in improving motor function and ADL. In these studies, it is shown that the neuromotor system can be activated correctly by providing a change in the sensory inputs to muscle and joint receptors as well as the skin receptors of patients' feet \cite{Kiper_2015}. Lim \cite{Lim_2019} report that a multi-sensory training program significantly improves proprioception and balance ability in patients; however, these types of studies are very limited.

The main goal of rehabilitation is the recovery of lost motor skills permanently and as quickly as possible. The effective way of training in motor learning can go through optimization of given tasks and feedback by variable sensory inputs such as pressure, vibration, and proprioception. These inputs not only facilitate motor learning but also develop compensatory mechanisms and strategies to overcome the loss of motor function resulting from a damaged neuromuscular system \cite{Sigrist_2013}. Feedback (visual, auditory, or tactile) is shown to improve complex motor learning. However, in daily life, multi-modal stimuli are more dominant than uni-modal stimuli since they provide convenience in ADL. Healthy individuals process stimuli in different modalities simultaneously. Multi-modal stimuli enable the learning of several aspects of a movement at the same time. Certain advantages of each modality are exploited, such as the ability of visualizations to show spatial aspects, and audio or tactile feedback to show temporal aspects. Moreover, it is reported that this sensory enhancement facilitates the transition between the senses (using other sense when one sense is inadequate). The researchers hypothesize that after training with multi-modal stimuli, sense processing would be active even when only uni-modal stimuli is present. For example, the learning process of motion detection tasks progressed positively even when auditory feedback was canceled after training with audio-visual feedback. This shows the importance of multi-modal training for complex motion recovery even at a further level \cite{Sigrist_2013, Pan_2019, Morone_2021}.

In recent years, rehabilitation strategies that include the active participation of patients and task-oriented exercises during rehabilitation sessions are carried out by virtual reality (VR). It is integrated to increase the attention and motivation of individuals with the desire for reward and success by giving continuous feedback \cite{Massetti_2016}. There are two different approaches to use VR in rehabilitation. The first one is called serious games, and they are specially designed for the rehabilitation robot/method. The second is called exercise games and refers to the use of games that already exist for entertainment purposes in rehabilitation \cite{Gon_2014}. With the integration of robotic rehabilitation in the VR environment, it is aimed to provide visual feedback about the results of the movements performed during the rehabilitation process, increase awareness about the quality of the movements, and accelerate motor learning \cite{Maggio_2019}. By including haptic cues during these applications, gait parameters such as symmetry, balance, and muscle activation patterns can be improved collectively. Thus, it is observed that the motor and cognitive status of patients can be improved at high rates by providing multi-sensory feedback and repetition of tasks with the robotic device, encouraging patients to actively participate in rehabilitation \cite{Maggio_2019, Feys_2019}.

A systematic review and meta-analysis are conducted to investigate the effectiveness of vibrotactile feedback (VF) on balance and gait rehabilitation. It shows that haptic feedback could represent a helpful intervention \cite{De_2021} which is generally divided into two types; kinesthetic and tactile. The former cue usually contains a sense of force and provides the user with a spatial frame of reference \cite{Verna_2020, Van_2017}, while the latter cue usually includes a sense of vibration, texture, or pressure. Such feedback can be delivered via existing interfaces, which offer to the user kinesthetic and tactile sensations \cite{Scotto_2020, Hocaoglu_2019}. When paired with other feedback, i.e., kinesthetic \cite{Afzal_2018} with visual \cite{Lee_2015}, the beneficial effects of VF on gait and balance parameters are found to be stronger. Because of these characteristics, VF can be employed as a supportive sensory stimulus in the context of a rehabilitation intervention focusing on sensorimotor integration. VF is also shown to be therapeutic for patients with neurological disorders \cite{Otis_2016, Carey_2016, Meyer_2016}.

The interdependence of sensory, cognitive, and motor processes, as well as the need for integrated training are increasingly employed in therapies to improve ADL, motor skills and balance, in the meantime to decrease fear of falling. In other words, APAs and CPAs (proactive and reactive control adjustments) can be retrained to improve QoL.

\subsection{Robot-Aided rehabilitation}\label{sec:Robot-Aided Rehabilitation}
Robot-assisted rehabilitation is continuously gaining prominence as it provides more efficient training and objective evaluation than conventional rehabilitation procedures \cite{Shakti_2018, Saglia_2009, Saglia_2010}. Furthermore, conventional methods require at least three physiotherapists to support the lower extremity and trunk of the patient manually. Another limitation is that the effectiveness of the rehabilitation during these practices depends on the personal knowledge and experience of the therapist. It is reported that the demand for physiotherapists is increasing continuously to match the number of patients due to the increase in the aging population worldwide. The use of robotic devices to address these challenges is encouraged to shift the adoption of rehabilitation clinics from conventional methods to robotic-assisted rehabilitation. Hence, high-quality therapy sessions can be achieved at a relatively low cost and with significantly less effort \cite{Krebs_2013, Kalita_2021, Daz_2011, Yurkewich_2015}.

Most neurological patients have a reduced range of motion (ROM) in their ankle as well as muscle strength in their bodies. In the literature, muscle-strengthening for ankle rehabilitation has been considered with three main phases; ROM, strength, and proprioceptive training, respectively, and they require different control methods. At the beginning of the rehabilitation, the patients have limited ankle mobility, thus passive ROM exercises are required (the patient is passive and the device is fully active) so that ankle ROM can be recovered via full assistance of the device. Then, active ROM exercise is utilized by reducing the level of assistance from device, and the patient should put effort to initiate the motion against the device. The second phase is about improving the ankle stability; as the patient's response improves, the resistance level of these strengthening activities increases. In most cases, they are unable to exert enough force to complete the exercises, so the robot should assist them. In the final phase, proprioceptive training should be performed to improve the balance control \cite{Val_2017, Bernhardt_2005, Teramae_2018}. However, due to their limited design configuration, i.e., low-weight bearing capacity or small end effector area, most rehabilitation platforms cannot be used in the balance rehabilitation; the detailed description is given in Section \ref{sec: Design of rehabilitation robots}. The control strategies in the stated phases are also illustrated in Table \ref{tab1}.

\begin{table}
\caption{Control Methods in Rehabilitation  \cite{Saglia_2010}}
\label{table}
\begin{tabular}{|p{100pt}|p{90pt}|p{90pt}|}
\hline
Exercise method& 
Patient mode& 
Control method\\
\hline
 ROM & Passive & Position Control \\ 
     & Active & Assistive Control  \\ \hline
 Strength Training  & Active- Isometric & Position Control  \\
     & Active- Isotonic & Admittance Control \\ \hline
 Proprioceptive Training  & Active & Hybrid Control  \\ 
\hline
\end{tabular}
\label{tab1}
\end{table}

Even though, robotic-assisted rehabilitation has a promising effect on the treatment itself, especially with the capability of being repetitive and task-specific, implementation of these potentials at the desired level is still challenging. There are numerous methods for control of these robots, yet control-related challenges are to be addressed \cite{Li_2017, Dong_2021, Marchal_2009, Jin_2018, Dzahir_2014}. Position control is one of the most used methods due to its simplicity. In this approach, the robot tries to follow a predefined motion trajectory \cite{Mohebbi_2020}. Furthermore, the position control is needed to conduct the movements in the first phase along a specific trajectory at a constant speed to improve the ROM capacity of the patient \cite{Ayas_2017, girone1999rutgers, Zhang_2016, Saglia_2013}. However, this method cannot be used in the other phases since when the patient applies more than the expected force/torque during rehabilitation, the position method cannot compensate for this disruptive effect, which leads to excessive position tracking errors. Despite this, it should be noted that most of the existing interaction controllers continue to use a position control scheme as an inner control loop, with a corresponding force/torque or motion outer loop applied to complete the interaction controller \cite{lu_2016, Song_2019}.

To eliminate the problem with position control, impedance control, which provides the desired dynamic interaction between the robot and its environment is proposed. The physical interaction is expressed as the dynamic relationship between the motion variables of the manipulators and the contact forces that need to be kept within a predetermined safe or acceptable position trajectory while the robot follows the desired motion trajectory \cite{Song_2019}. Admittance control uses force as an input and displacement as an output while the robot follows predetermined force. These methods are used in the strength phase to assist/resist the patients \cite{Saglia_2013, Jamwal_2016, Ibarra_2015, Sun_2015}. However, these methods require force/torque sensor to measure the force exerted by patients, which adds extra complexity, i.e., sensor dynamics, cost, and computational load. Moreover, the estimation of impedance parameters peculiar to the individual is another challenge \cite{Codourey_2016}.

The dynamic model of robots is usually composed of nonlinear functions of the state variables (joints' positions and velocities), especially in parallel structures. This characteristic of the dynamic model makes the closed-loop control system nonlinear and difficult to solve. The computed torque control (CTC) method requires a good knowledge of the robot dynamics since the dynamic model of the manipulator is used in the loop. Even though, CTC can compensate for non-linearity since dynamic equations are solved in real-time, it creates a computational burden \cite{Codourey_2016, Tsoi_2009, Asgari_2015}.  

The above-mentioned issues are addressed by using the parameter estimation control method. With this method, the simplification of complex mathematical dynamic equations and modeling of unmodeled noise signals are done and a new physical model with functional properties can be obtained with less computation time and acceptable control performance. Furthermore, modeling can be done online or offline. In the offline method, if the estimated model is not highly accurate, it cannot be able to correctly distinguish between responses caused by known and unknown input signals. In the online method, parameter estimation is done simultaneously within the process, causing a delay in the system response \cite{Song_2019, Gao_2014, Wolbrecht_2008}. The fuzzy logic method is also used in the control of rehabilitation robots. Furthermore, it provides good performance for the control of the nonlinear system. Fuzzy logic is similar to human thought systematic than traditional logic systems. Basically, it tries to capture the approximate, imprecise nature of the real world. It has three steps; fuzzification, linguistic rules, and defuzzification. In the fuzzification step, input signals are converted into fuzzy sets with some degree of membership (range from 0 to 1). The main part of fuzzy logic control is controlling the robot using a set of linguistic control rules associated with binary concepts such as fuzzy inference and computational inference rules. In the defuzzification, fuzzy truth values are converted into output decision values. However, like parameter estimation, it is not robust against unexpected situations during real-time control since its rules are determined previously \cite{Sharma_2020, Karasakal_2005, Lamamra_2020}.

In the above-mentioned methods, the patient passively follows the movements of the robot, which follows the previously specified position or force/torque references \cite{Patarinski_1993, Chiaverini_1993}. Recent studies in robotic therapy show that continuous passive training therapy does not significantly improve motor function. Active participation of patients is considered to be a major factor contributing to the neural plasticity and motor recovery \cite{Teramae_2018, Keller_2013}. One of the most commonly adopted assistance strategies, the "Assist-as-Needed (AAN)" paradigm, offers the mode of necessary active assistance to stimulate neuroplasticity. The basic principle in AAN is to provide physical assistance only when needed by the patient. If a patient performs a task flawlessly, the robotic assistance is withdrawn. However, if the patient has difficulty or cannot complete the task, the robot provides as much support as the patient needs to perform the task \cite{Pehlivan_2017}. In other words, AAN is a strategy of regulating auxiliary forces/torques or task difficulty according to patients' disability level or performance in training tasks. There is strong evidence that active participation induces neural plasticity, and therefore controllers should intervene minimally to promote participation and recovery. In addition, upper extremity rehabilitation using the AAN paradigm is shown to be the most promising technique for promoting recovery \cite{Luo_2019}.

New technological devices and methods are being developed to increase active patient participation by integrating multi-sensory information and allowing AAN paradigm. However, the implementation of AAN paradigm is still challenging, since determining the level of assistance according to the patient progress is not to be addressed sufficiently.

\subsection{Design of existing rehabilitation robots}\label{sec: Design of rehabilitation robots}
Considering the design and development of robotic devices for lower extremity rehabilitation, there are mainly three types of system, i.e., wearable exoskeleton system, ankle platform and balance rehabilitation platform are proposed \cite{Deng_2018, Daz_2011, Ersoy_2021}. Several examples of exoskeleton devices aim to increase the capacity of the lower limbs or reducing user effort \cite{Dollar_2008, Ferris_2006, Mooney_2014, Pratt_2004}. Systems such as AKROD \cite{Weinberg_2007}, BioMot project \cite{Bacek_2017}, KNEXO \cite{Beyl_2009}, Lokomat \cite{Jezernik_2003}, LOPES \cite{Van_2006}, MIRAD project \cite{Mirad}, and REX \cite{RexBionics} are used to support individuals with muscle weakness in ADL. However, since the mechanisms underlying human movements and how the designed devices should interact with humans are not fully understood, there is no device that can effectively improve the user's performance. The mentioned systems have difficulties in use because they have a rigid, bulky structure, and uncomfortable interfaces, restrict biological joints and are misaligned with natural joints. In addition, if the exoskeleton does not have enough degree of freedom (DoF) to work in harmony with human joints, it exerts a residual force on the human limb due to axial misalignment, and this may cause long-term injuries, as well as discomfort \cite{Schiele_2008, Schiele_2009}.

Platform-based robots are grounded and have movable end effector as a rehabilitation platform with one or more DoF. These types of systems employed for the ankle joint focus only on improving the ROM of the joint rather than improving the balance of the patient \cite{Daz_2011, Morris_2011}. Most of the platforms for the ankle joint are in parallel structure, which provides sufficiently high-torque for plantar flexion/dorsiflexion, inversion/eversion and adduction/abduction movements of the ankle \cite{Rastegarpanah_2016,Chablat_1998}. There are also systems designed to be serially connected to each other with motor-operated joints \cite{Saglia_2019}. Although serial manipulators are easier to model, using a parallel manipulator in such an application provides advantages in terms of achieving high load-carrying capability, better dynamic performance, and precise positioning.
Rutgers ankle is a Stewart platform-type haptic interface that provides 6 DoF resistance forces to the patient's foot in response to VR-based exercises \cite{Girone_2001}. Various clinical studies are conducted with this device showing improvement in patient strength and endurance measurements \cite{Deuschl_2020, Deutsch_2001b, Cioi_2011, Girone_2000}. In RePAiR, a 1-DoF robotic platform allows dorsiflexion and plantar flexion movements to patients after stroke. It is demonstrated that the device provides benefits in increasing the muscle strength of the patients, improving the motor control, sensory-motor coordination of the patients, and accordingly, the walking patterns \cite{Gon_2014}. Only a few of the manipulators developed for ankle rehabilitation are commercialized \cite{Saglia_2013, OptiFlex, Breva}. These platforms are widely used to strengthen ankle joint movements and improve ankle proprioception. The end effector of the above-mentioned manipulators are only one foot large and have a low weight-bearing capacity; therefore, they cannot be used for balance rehabilitation after ankle treatment has been completed. In addition, the end effector of the proposed systems are not endowed with a sensor; therefore, pressure change measurement on the sole and sensory input under the foot cannot be performed during the ROM rehabilitation.

Posturography, measurement of CoM and balance variables, are tested through static or dynamic techniques for balance evaluation and rehabilitation \cite{Prosperini_2013, Park_2014}. Training with static balance platforms is said to be helpful in controlling pressure distribution in patients over time. Static posturography is usually done by using Wii fit \cite{Wii} board or commercially available force platforms \cite{Bertec, Kistler, HURLabs, AMTI}. Studies evaluated the Wii exercise experience of the patients and their physiotherapists, are stated that Wii exercise is an amusing and challenging way to improve balance impairment since VR provides continuous feedback \cite{Prosperini_2013, Plow_2014}. Another study was conducted for 6-week with commercially available force platform and virtual reality and the result shows that the CoM points deviation decreased. However, when compared to static situations, rehabilitation under dynamic conditions contributes more to the improvement of balance disorders and motor skills \cite{Prosperini_2013}.

Dynamic platforms require instantaneous dynamic movements, forcing patients to adjust their balance during perturbation \cite{Prosperini_2013}. A study shows that dynamic strength platforms are more helpful in restoring postural stability than conventional therapy \cite{Saglia_2019}. Wooden balance platforms are one of the simplest examples, but they do not provide any quantitative measurement \cite{Balance}. The Bobo Balance platform provides force measurement on top of wooden platforms. However, it is difficult to use for people with severe loss of balance since there is no external support environment or mechanisms for patient while standing \cite{bobo-balance}. gePRO \cite{Vertigomedprodotti}, BackinAction \cite{RiabloEuleria}, Balanceback \cite{Balanceback}, Proprio \cite{Proprio}, and Huber \cite{HUBER} platforms have 2-DoF (roll and pitch) for balance rehabilitation. These systems deliberately put patients in an unbalanced state while patient following the VR game, thus assessing their balance status based on CoM position. During the dynamic rehabilitation process, the angle of the platform can be controlled, and patients are requested to maintain their CoM and posture. Even though ROM can be covered fully by gradually changing the angular position, these systems cannot provide AAN paradigm (even if a patient performance improves, the system's level of assistance remains unchanged) \cite{Marchal_2009, Kang_2015}. Furthermore, weight transfer in balance training cannot be provided during PBT due to the lack of sensory input to the foot with these devices. Additionally, the VR environment is the only sensory input proposed with these devices; however, it has been stated in the literature that multi-sensory input is more powerful for mimicking and enhancing ADL \cite{Sigrist_2013}.

Investigation of APAs and CPAs, which is necessary for proactive and reactive balance control, is claimed to disclose essential aspects of postural control and history of falls. The studies show severe APAs and CPAs deficiencies after evaluating the patient's condition and muscular activity; however, retraining APAs and CPAs is possible through PBT programs. Despite the benefits of these studies, to our best of knowledge, there is no study in the literature that assesses reaction time, CoM, CoP and the ability to control APAs and CPAs on ankle, balance, and stepping rehabilitation simultaneously. On top of that, the effects of multi-sensory inputs and cognitive control strategies suitable for the use of patients with different severity levels have not been evaluated and utilized. 

\section{Results}
The ability of balance is a multivariate concept that integrates both motor and sensory components. Postural control during standing and walking requires multi-modal sensory feedback, e.g., visual, vestibular, and proprioceptive feedback \cite{Horak_2006}. Sensory inputs have contributions depending on the environment and the motor task performed by the individual, and patients with neurological damage live difficulties in weighing and utilizing sensory inputs \cite{Costantino_2017,Negahban_2011}. 

Rehabilitation of balance should be performed with the integration of ankle-foot, balance, and stepping phases in order to enhance activity-based neuroplasticity. Therefore in this study, we propose an I-BaR framework that adopts AAN paradigm for balance analysis, rehabilitation and assistance with multi-modal feedforward and feedback signals.

\begin{figure}[t]
\centering
\includegraphics[width=\textwidth]{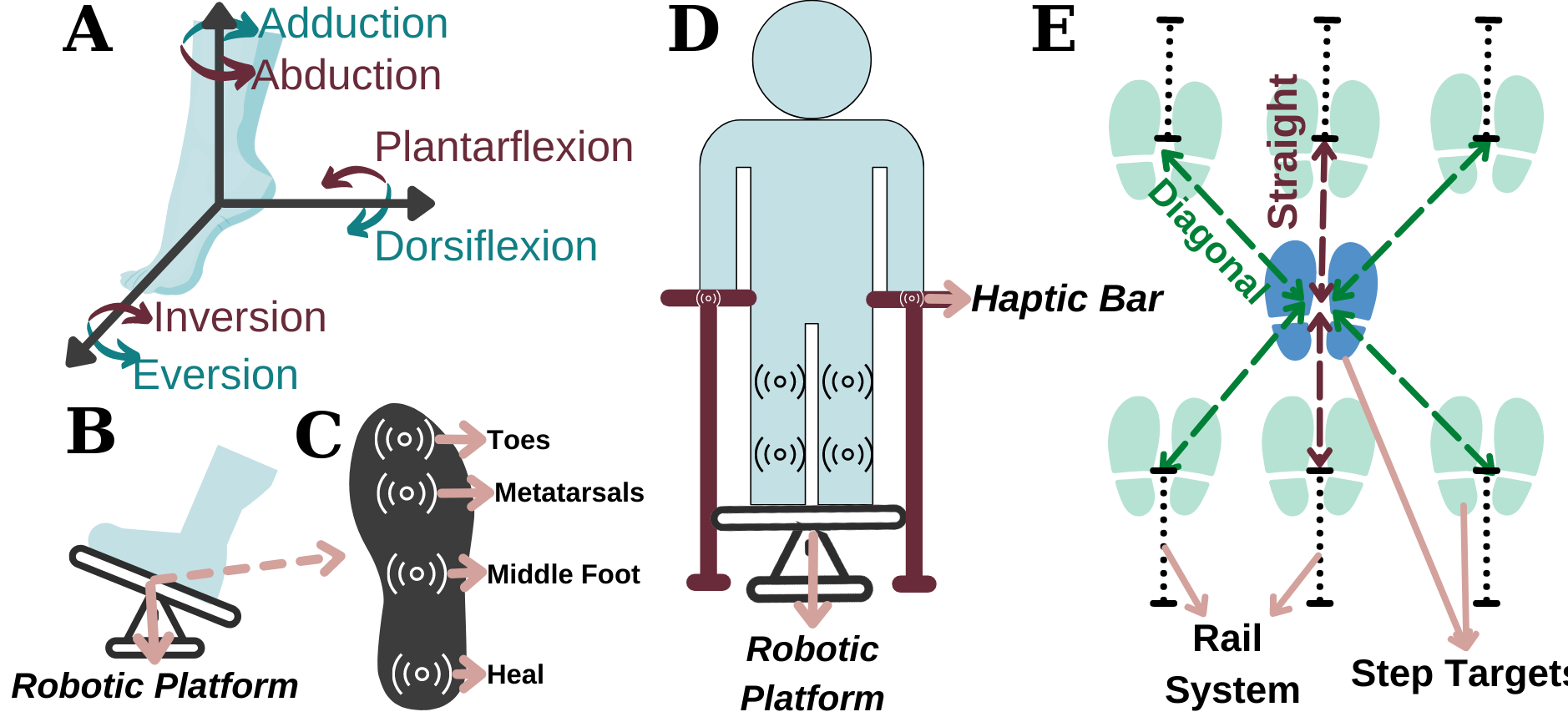}
\caption{Representation of (A) the ankle DoF, (B) ankle-foot rehabilitation, (C) sole of the foot, (D) balance, and (E) stepping rehabilitation}
\label{Ibar1}
\end{figure}

It is known that the sensory input training can positively affect motor control during balance rehabilitation \cite{Carey_2016, Bottaro_2008, Rossignol_2006, Laaksonen_2012}. Multi-modal information provides certain advantages in terms of effective and realistic training to mimic ADL. For instance, the ability of visualization shows spatial aspects, while audio/tactile feedback allows temporal aspects. Studies on the neurophysiology of somatosensory information processing emphasize that multiple cortical and subcortical brain (CNS and PNS) structures take an active role in sensory discrimination tests \cite{Sigrist_2013, Lee_2018, Pan_2019}. Moreover, a stimulators should be placed on the skin of the patient so that muscles with low activity in sEMG measurement can be triggered. VR environment should be employed to improve the effectiveness of the training and patient engagement while serving as visual and auditory feedback. VR tasks should be designed similarly to the activities that the patient has difficulty and their difficulty levels should be adjustable.

\begin{figure}[t]
\centering
\includegraphics[width=\textwidth]{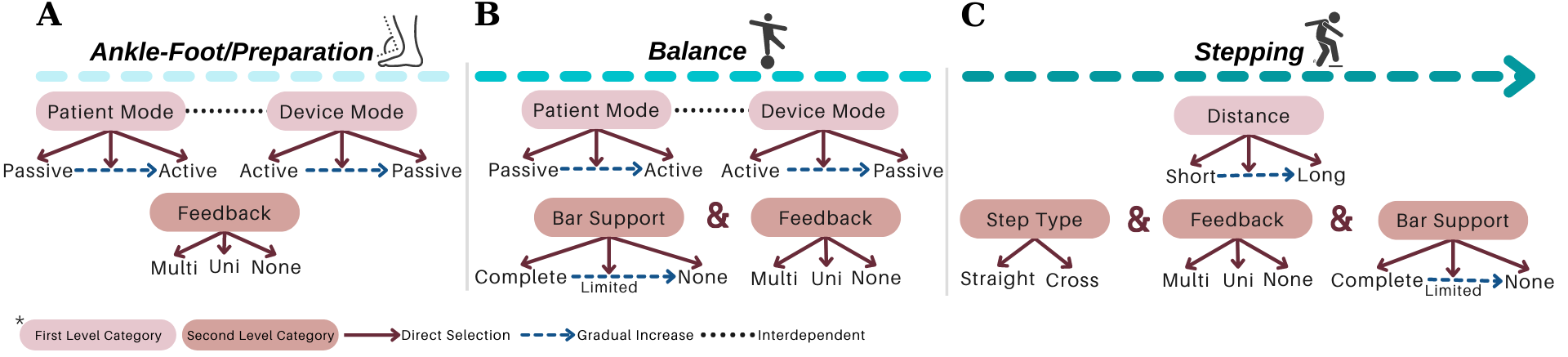}
\caption{Protocol of I-BaR: (A) Ankle-Foot/Preparation, (B) Balance, and (C) Stepping Phases}
\label{Ibar2}
\end{figure}

I-BaR framework proposes a personalized approach that allows patient-specific difficulty levels in three main phases of rehabilitation (see Figure \ref{Ibar2}). In ankle-foot phase, after the mode of the patient (see Figure \ref{Ibar2}A) is determined and accordingly the selected robotic device should be controlled with the AAN paradigm, feedback mode can be selected according to the patient's needs. Similarly, for the balance phase (see Figure \ref{Ibar2}B) personalized rehabilitation can be offered according to the patient's capability by selecting different combinations from the bar support and feedback mode section. Moreover, after the distance is determined in stepping rehabilitation (see Figure \ref{Ibar2}C), rehabilitation can be done with the combinations of the feedback mode and step type sections mode. In this way, rehabilitation can be provided to the patient at different severity levels in the areas they completely lack. Since each option will be increased gradually throughout the process, he/she can continue their treatment without difficulty.

\subsection{Ankle-Foot/Preparation Phase}
The first phase is ankle-foot/preparation rehabilitation to improve lower extremity muscles while the patient is sitting due to their ankle instability and low muscle activation. It aims to prepare the sensorimotor system for motor function, which is essential when there is minimal or no voluntary motor/muscle function in ankle-foot. The selected robotic platform should be kept in ankle ROM, shown in Figure \ref{Ibar1}A  with 0-2$0^o$ dorsiflexion, 0-5$0^o$ plantarflexion, 0-1$0^o$ adduction, 0-$5^o$ abduction, 0-2$0^o$ eversion and 0-3$5^o$ inversion limits \cite{Hasan_2020}. The system needs to ensure that each patient can perform exercises at their specific ankle ROM limits. Upper limits must be determined according to the impairment levels of the patient and it should be gradually increased based on their performance. Such limits can be identified with the use of a robotic platform with a predetermined force/torque based on an individual's parameters, i.e., foot mass, size and inertial parameters. This torque should have the maximum magnitude to be able to move the individual's foot. Patient can exert reaction force/torque due to spasticity (involuntary muscle contraction) and RoM limits during robotic platform induced motion. Thus, once platform cannot move the foot any further, this angle is considered as patient's spasticity level. Up until this time rehabilitation is performed with patient passive mode; however as patient progresses in the therapy, his/her contribution to the motion would also be requested at this phase (patient active mode). For instance, moving the ankle to $5^o$ of dorsiflexion with patient active contribution while the platform supports and this support can be regulated according to his/her progress.  

Moreover, 43\% of patients have decreased sensation of their soles after stroke and it is also been indicated that providing sensory feedback to the sole of the foot (according to their vibration sensing level at sub-threshold) drastically lowers postural sway during standing in older adults, diabetes, and stroke patients \cite{Chien_2017}. Therefore, a force and haptic feedback sensor should be placed in at least four regions, namely toes, metatarsals, middle foot, and heel, as they are more intense regions \cite{Sport_Education_2002} (see Figure\ref{Ibar1}B for feet placement and see Figure\ref{Ibar1}C for specified regions). With the force sensor, the pressure applied by the patient to the each region  on the sole can be measured and if the pressure distribution exceeds the able-bodied pressure distribution value in one of these regions, he/she can be stimulated that region via haptic feedback based on the individually identified sensing threshold. Accordingly, the patient can be aware of their amount of weight-bearing and sole pressure, thus they can be trained to transfer their weight correctly for fall prevention. With this aim, a personalized preparation therapy program is proposed as an essential stage for I-BaR, which provides multi-modal feedback/feedforward signal and determines the somatosensory level of patient for the therapist to plan and decide on the parameters of the next phase.

\subsection{Balance Phase} 
The second phase is balance rehabilitation to improve postural adjustments while the patient is standing on the platform. Although the patient pass to this phase after the preparation, they may have difficulty in balancing their posture without physical assistance. Therefore, a haptic bar should be installed on the selected rehabilitation system (see Figure\ref{Ibar1}D). This haptic bar should at least include two sensors, force and haptic sensors, to quantitatively measure applied force and give haptic feedback accordingly. In the case of a patient with higher balance loss, the phase should be executed with full assistance of haptic bars. Throughout the rehabilitation as the patient's postural control improves, an upper extremity support limit value should be established for each patient. When the patient goes over the limit, haptic feedback is applied to warn them to reduce the upper extremity support and help them to control the posture by using the lower extremity rather than relying on the arms.

As the core part of balance rehabilitation, this phase should be utilized to increase sensory and motor integration during task-specific activities. Patient-specific information, such as ROM and proprioception, obtained from the preparation phase determine the required parameters of the therapy at this phase. This rehabilitation can also increase somatosensation during task-specific activity. Within the scope of these interventions, the patient's balance in the mediolateral and anteroposterior directions can be disrupted by giving sudden and fast perturbations with rehabilitation robot. In PBT, to train balance reactions, perturbations in different directions and amplitudes are applied, and various scenarios are used, including simultaneous cognitive and motor activities. The objective of this is to decrease the reaction time, and deviation in CoM and also to improve pressure distribution as well as APAs and CPAs \cite{Mansfield_2015, Jagdhane_2016}. As in the preparation phase, the AAN control paradigm should be implemented on the Robotic platform system during PBT so that the patient can engage with the dynamics of the platform independently and actively within the limits of their capacity. 

\subsection{Stepping Phase}
The third phase is stepping rehabilitation to improve walking parameters, i.e., the stance and swing phase's temporal and spatial parameters. While step speed is increasing, improving the stepping accuracy is another target in this phase. It should include assessment and training of stepping to various target distances with robotic help. Step-taking activities is performed to assess the effectiveness of the previous two phases by analyzing the relationship between muscle activity, postural control, movement speed, and accuracy as well as to train walking parameter with Fitts' law. Particularly, at stepping long distances the patient realize the movement faster with their own compensation strategies which reduces the quality of motion, whereas at stepping short distances their motion is slower, controlled and precise. Current step-taking assessments are based on manual change of the target distance and reaction time measurements (detected by a switch button according to the time between target point and initial position force detection) \cite{Yamada_2021,Aloraini_2020,Aloraini_2019}. Since force measurements and haptic sensors are not implemented at these target points, a kinetic evaluation and improvement in sensory input under the foot cannot be provided. Therefore, in the stepping phase, the distance of the target points should be automatically adjustable within the system to provide perturbation/moving target in x-y plane and these target points should include force and haptic sensors to calculate GRF and increase sensation under the sole, respectively (see Figure \ref{Ibar1}E).

\section{Discussion}\label{sec:Discussion}
Rehabilitation helps with the physiological and functional recovery of motor and sensory skills that have been lost \cite{Riemenschneider_2018}. Recently, it has been stated that investigating the APAs and CPAs, which are required for proactive and reactive balance control, may reveal critical information about postural control and fall history. Early muscle activation and decreased CoM deviation provided vital evidence that retraining of APAs and CPAs is achievable for stroke and elderly population. The literature provides the fundamentals of the development of PBT to enhance postural control and balance in patients to prevent fall. The therapy procedure used for large body structures, such as the lower extremities and trunk, requires a lot of physical effort for therapists and patients, and the effectiveness of rehabilitation depends on the therapist's personal knowledge and experience \cite{Kalita_2021}. Robotic rehabilitation studies are increased significantly in order to overcome these difficulties since it is recognized as efficient and provide precise assessment of kinematic and dynamic parameters as well as objective evaluations that enhance the treatments. Yet, there are still challenges to be addressed, such as limited accessibility of present robotic devices due to cost, the training of all patients at the same level of difficulty, and insufficient personalized approach \cite{Saglia_2009, Ilett_2016}. 

To the best of our knowledge, the aforementioned limitations have been addressed in a separate manner. For instance, in ankle rehabilitation, robotic platforms are frequently used for patients with higher severity levels to improve ankle ROM, strength and proprioceptive. Although these studies are sufficient to improve ankle instability, PBT is still required for these individuals to improve their ADL. Ankle rehabilitation robots cannot be used in balance rehabilitation because they only have the surface area to fit one foot and have a low weight capacity compared to the whole body weight.

Moreover, in balance rehabilitation, treatment is implemented with either a static or dynamic system. Although static balance systems are useful in evaluating the patient's balance level, dynamic anteroposterior and mediolateral perturbations are required to improve plasticity in patients. Furthermore, in the dynamic balance systems, due to the lack of sensors on the end effector, these systems cannot train patients to control their weight transfer to avoid falls. In particular, the pressure change in certain regions of the foot (at least four major parts heal, middle foot, metatarsals, and toes) cannot be measured and corresponding vibrotactile feedback cannot be applied under the foot. Accordingly, such systems are not able to improve decreased sensation on the sole.

As the third main treatment, stepping rehabilitation is implemented to improve walking parameters, stepping accuracy and stepping ability to various target distances. Fitts' law is used to explore the role of motor planning processes. According to the law, the target's width and distance are correlated and will determine how long it takes to go quickly to a particular region. Furthermore, this law is implemented on the CSRT test to analyze the time to complete the foot-reaching task and improve the walking parameters. Various target points are placed on the environment, and the subject must step on the target accordingly to the given feedback. These target points just include a switch to determine reaction time. In other words, they are unable to assess GRF for haptic stimuli and train the patient to control pressure distribution. 

It is essential to associate the aforementioned rehabilitation method with multi-modal feedback since it enables the learning of several aspects of a movement simultaneously and mimics daily life, e.g., visual, pressure, audio, vibration, and proprioception. These feedback information not only promotes the development of plasticity but also provides compensation for the loss of motor function caused due to a compromised neuromuscular system. However, to the best of our knowledge, there is no study to prove the importance of multi-modal feedback in the improvement of postural adjustments and reaction time. 

The three-phase framework, I-BaR, is proposed to address the aforementioned drawbacks of current rehabilitation approaches. First of all, all these three separate treatments should be implemented as a whole in a single system to improve foot/ankle functionality and postural adjustments (APAs and CPAs) for achieving high-quality walking based on objective assessment.  Moreover, in all these phases, using the multi-modal feedback for improving sensory weighting skills, the AAN paradigm for modulating the level of assistance according to patients' progress and mimicking the situations that the patient encounters in ADL should be implemented as a whole for an effective assessment and rehabilitation to achieve gradual independence eventually. Therefore, the I-BaR framework offers an effective solution as a whole with these properties to achieve personalized balance rehabilitation for different disability levels. 

In this context, the requirements to be considered in order to implement the I-BaR framework and to ensure the development of the patients can be divided into two main groups and they are summarized in Table \ref{ta} and \ref{ta2}.

Physical system requirements;
\begin{itemize}
    \item At least 3-DoF (roll (rotation in x), pitch (rotation  in y), and (elevation in height (translational in z)) to mimic ADL
    \item System should be able to determine applied force/torque based on an individual’s parameters (Patient's ROM, ankle torque) to keep the treatment at the limit that the patient can have difficulty continuously
    \item AAN paradigm implementation on control structure to encourage active participation of the patient
    \item System's end effector should include at least four force and haptic sensors on each foot (should cover at least the toes, metatarsals, middle foot, and heel) to quantitatively measure applied force and give haptic feedback accordingly
    \item Haptic bar should be installed on the system to provide physical assistance on balance and stepping rehabilitation
    \item Haptic bar should be include force and haptic sensors to measure upper extremity support  and this support should be reduced by haptic feedback (if the patient depends to upper extremity to walk or control the posture) 
    \item Target point on the stepping rehabilitation should be automatically adjustable to provide perturbation/moving target in x-y plane
    \item Target point on the stepping rehabilitation should be include force and haptic sensor to quantitatively measure walking parameter (GRF on stepping) and increase sole sensation 
\end{itemize}

Rehabilitation requirements;
\begin{itemize}
    \item Muscle which has low activity should be triggered with haptic feedback to initiate the motion during all phases
    \item Each patient should be perform the exercises at their specific ankle ROM and torque/force to maintain the patient safety
    \item Patient’s balance in the mediolateral and anteroposterior directions should be disrupted by large and sudden perturbation to improve plasticity and balance reactions (neuromuscular responses)
    \item According to the patient's at sub-threshold vibration sensing level, haptic feedback should be provided to improve decreased sole sensation
    \item Multi-modal feedback (Haptic, audio, visual, physical (anteroposterior and mediolateral) perturbation) should be used to improve sensory-integrating weighting skills  
    \item Ankle-foot, balance, and stepping rehabilitation should be performed as a whole, starting from the necessary phase according to the patient's initial physical condition, to achieve gradual independence.
\end{itemize}

\begin{table}
\caption{I-BaR Requirements (Limit reference for ROM:\cite{Hasan_2020}, PBT:\cite{Freyler_2015} and Stepping:\cite{krebs_2001}  )}
\label{ta}
\begin{tabular}{|p{50pt}|p{90pt}|p{180pt}|}
\hline Exercise method& Limit & Personalization\\\hline
ROM & 2$0^o$ dorsiflexion, 5$0^o$ plantarflexion, 1$0^o$ adduction, $5^o$ abduction, 2$0^o$ eversion and 3$5^o$ inversion & Each patient maximum ROM limit should be determined accordingly to torque/force applied on the system (based on the sEMG and force sensor measurement)\\ \hline
PBT & The displacement of CoM and perturbation velocity should be 2-3 cm 0.11-0.18 m/s, respectively & Each patient perturbation speed should be arranged accordingly to the patient postural control level (based on the sEMG, force sensor measurement, CoM deviation and haptic bar/upper extremity support) \\ \hline
Stepping target perturbation  & Healthy adult walking speed 1.2–1.3 m/s  & Target position should be determined accordingly the second phase response (based on the sEMG, force sensor measurement, CoM deviation and haptic bar/upper extremity support)\\ \hline
\end{tabular}
\end{table}

\begin{table}
\caption{Feedback/Feedforward Requirement}
\label{ta2}
\begin{tabular}{|p{60pt}|p{260pt}|}
\hline
Sensor/Method& Limit \\ \hline
Haptic Sensor & Threshold value should be determined accordingly to individuals EMG and weight transfer capacity \\ \hline
Force Sensor & Weight capacity should be at least 100 kg to cover so that 95\% of the user population would be compensated.\\ \hline
Position Control & Should be used to prepare the sensorimotor system for motor function when there is minimal or no voluntary motor/muscle function
in ankle-foot. \\\hline
Impedance Control & Should be used to adjust patient’s spasticity level (ability of force application) \\\hline
AAN Paradigm & Should be implemented in the control structure to provide physical assistance only when needed by the patient and thus to stimulate neuroplasticity \\\hline
Feedback type & Multi-modal sensory feedback (Audio, haptic visual, vestibular, and proprioceptive) should be implemented on the rehabilitation procedure to integrates both motor and sensory components as whole and thus improve ADL. \\\hline
\end{tabular}
\end{table}

\section{Conclusions}\label{sec:Conclusion}
The interdependence of sensory, cognitive, and motor processes, as well as the need for integrated training, are increasingly being employed in therapies to increase motor skills and balance, and decrease the fear of falling. According to these approaches, new technological devices and paradigms must be developed to promote the participation of a paradigm from the CNS to the PNS. 

In this context, we explain the necessity of the proposed I-BaR framework, which  includes:
\begin{itemize}
 \item{Ankle-foot Rehabilitation:} ankle-foot muscle activation, sole, joint, and movement sensations are developed while sitting,
 \item{Balance Rehabilitation:} sensory weighting skills are developed from motor learning by using multi-sensory input during PBT and help gradual independent,
 \item{Stepping Rehabilitation:} walking parameters are improved during step-taking activities to target points with adjustable distance.
\end{itemize}

In all the phases mentioned above, using the multi-modal feedback for improving sensory weighting skills, AAN paradigm for modulating the level of assistance according to patients' progress and mimicking the situations that the patient encounters in ADL should be implemented as a whole for an effective assessment and rehabilitation to achieve gradual independence eventually. 

As a future work, the design and construction of the required robotic platform are in our ongoing study for validating and evaluating this framework on patients with neurological disorders.

\section*{Declarations}
\subsection*{Funding}
This research is partially supported by The Scientific and Technological Research Council of Türkiye (TUBITAK) under Grant 122E246. The funders had no role in study design, data collection and analysis, decision to publish, or preparation of the manuscript.
\subsection*{Conflict of interest/Competing interests}
  The authors declare that they have no competing interests.
\subsection*{Ethics approval}
Not applicable - This study does not conducted on human subject experiment.
\subsection*{Consent to participate}
Not applicable - This study does not conducted on human subject experiment.
\subsection*{Consent for publication}
Not applicable.
\subsection*{Availability of data and materials}
Not applicable.
\subsection*{Code availability}
Not applicable.
\subsection*{Authors' contributions}
TE, PK, EH, and RU conceived the I-BaR framework. TE and PK were responsible for the literature survey of the whole system. PK, EH, and RU supervised the study. All authors contributed to the manuscript revision, read, and approved the submitted version.

\noindent

\bigskip

\bibstyle{sn-nature.bst}
\bibliography{maintext}


\begin{thebibliography}{174}
\ifx \bisbn   \undefined \def \bisbn  #1{ISBN #1}\fi
\ifx \binits  \undefined \def \binits#1{#1}\fi
\ifx \bauthor  \undefined \def \bauthor#1{#1}\fi
\ifx \batitle  \undefined \def \batitle#1{#1}\fi
\ifx \bjtitle  \undefined \def \bjtitle#1{#1}\fi
\ifx \bvolume  \undefined \def \bvolume#1{\textbf{#1}}\fi
\ifx \byear  \undefined \def \byear#1{#1}\fi
\ifx \bissue  \undefined \def \bissue#1{#1}\fi
\ifx \bfpage  \undefined \def \bfpage#1{#1}\fi
\ifx \blpage  \undefined \def \blpage #1{#1}\fi
\ifx \burl  \undefined \def \burl#1{\textsf{#1}}\fi
\ifx \doiurl  \undefined \def \doiurl#1{\url{https://doi.org/#1}}\fi
\ifx \betal  \undefined \def \betal{\textit{et al.}}\fi
\ifx \binstitute  \undefined \def \binstitute#1{#1}\fi
\ifx \binstitutionaled  \undefined \def \binstitutionaled#1{#1}\fi
\ifx \bctitle  \undefined \def \bctitle#1{#1}\fi
\ifx \beditor  \undefined \def \beditor#1{#1}\fi
\ifx \bpublisher  \undefined \def \bpublisher#1{#1}\fi
\ifx \bbtitle  \undefined \def \bbtitle#1{#1}\fi
\ifx \bedition  \undefined \def \bedition#1{#1}\fi
\ifx \bseriesno  \undefined \def \bseriesno#1{#1}\fi
\ifx \blocation  \undefined \def \blocation#1{#1}\fi
\ifx \bsertitle  \undefined \def \bsertitle#1{#1}\fi
\ifx \bsnm \undefined \def \bsnm#1{#1}\fi
\ifx \bsuffix \undefined \def \bsuffix#1{#1}\fi
\ifx \bparticle \undefined \def \bparticle#1{#1}\fi
\ifx \barticle \undefined \def \barticle#1{#1}\fi
\bibcommenthead
\ifx \bconfdate \undefined \def \bconfdate #1{#1}\fi
\ifx \botherref \undefined \def \botherref #1{#1}\fi
\ifx \url \undefined \def \url#1{\textsf{#1}}\fi
\ifx \bchapter \undefined \def \bchapter#1{#1}\fi
\ifx \bbook \undefined \def \bbook#1{#1}\fi
\ifx \bcomment \undefined \def \bcomment#1{#1}\fi
\ifx \oauthor \undefined \def \oauthor#1{#1}\fi
\ifx \citeauthoryear \undefined \def \citeauthoryear#1{#1}\fi
\ifx \endbibitem  \undefined \def \endbibitem {}\fi
\ifx \bconflocation  \undefined \def \bconflocation#1{#1}\fi
\ifx \arxivurl  \undefined \def \arxivurl#1{\textsf{#1}}\fi
\csname PreBibitemsHook\endcsname

\bibitem[\protect\citeauthoryear{}{2022}]{Who_2022}
\begin{botherref}
WHO.
\url{https://www.who.int/publications/i/item/9789240051157}.
Accessed: Nov. 28, 2022
(2022)
\end{botherref}
\endbibitem

\bibitem[\protect\citeauthoryear{Fineberg et~al.}{2013}]{Fineberg_2013}
\begin{barticle}
\bauthor{\bsnm{Fineberg}, \binits{N.A.}},
\bauthor{\bsnm{Haddad}, \binits{P.M.}},
\bauthor{\bsnm{Carpenter}, \binits{L.}},
\bauthor{\bsnm{Gannon}, \binits{B.}},
\bauthor{\bsnm{Sharpe}, \binits{R.}},
\bauthor{\bsnm{Young}, \binits{A.H.}},
\bauthor{\bsnm{Joyce}, \binits{E.}},
\bauthor{\bsnm{Rowe}, \binits{J.}},
\bauthor{\bsnm{Wellsted}, \binits{D.}},
\bauthor{\bsnm{Nutt}, \binits{D.J.}},
\bauthor{\bsnm{al.}}:
\batitle{The size, burden and cost of disorders of the brain in the uk}.
\bjtitle{Journal of Psychopharmacology}
\bvolume{27}(\bissue{9}),
\bfpage{761}--\blpage{770}
(\byear{2013})
\doiurl{10.1177/0269881113495118}
\end{barticle}
\endbibitem

\bibitem[\protect\citeauthoryear{Pehlivan et~al.}{2016}]{Pehlivan_2016}
\begin{barticle}
\bauthor{\bsnm{Pehlivan}, \binits{A.U.}},
\bauthor{\bsnm{Losey}, \binits{D.P.}},
\bauthor{\bsnm{Omalley}, \binits{M.K.}}:
\batitle{Minimal assist-as-needed controller for upper limb robotic
  rehabilitation}.
\bjtitle{IEEE Transactions on Robotics}
\bvolume{32}(\bissue{1}),
\bfpage{113}--\blpage{124}
(\byear{2016})
\doiurl{10.1109/TRO.2015.2503726}
\end{barticle}
\endbibitem

\bibitem[\protect\citeauthoryear{Gaskin et~al.}{2017}]{Gaskin_2017}
\begin{barticle}
\bauthor{\bsnm{Gaskin}, \binits{J.}},
\bauthor{\bsnm{Gomes}, \binits{J.}},
\bauthor{\bsnm{Darshan}, \binits{S.}},
\bauthor{\bsnm{Krewski}, \binits{D.}}:
\batitle{Burden of neurological conditions in canada}.
\bjtitle{NeuroToxicology}
\bvolume{61},
\bfpage{2}--\blpage{10}
(\byear{2017})
\doiurl{10.1016/j.neuro.2016.05.001}
\end{barticle}
\endbibitem

\bibitem[\protect\citeauthoryear{Mehravar et~al.}{2015}]{Mehravar_2015}
\begin{barticle}
\bauthor{\bsnm{Mehravar}, \binits{M.}},
\bauthor{\bsnm{Yadollah-Pour}, \binits{N.}},
\bauthor{\bsnm{Tajali}, \binits{S.}},
\bauthor{\bsnm{Shaterzadeh-Yazdi}, \binits{M.J.}},
\bauthor{\bsnm{Majdinasab}, \binits{N.}}:
\batitle{The role of anticipatory postural adjustments and compensatory control
  of posture in balance control of patients with multiple sclerosis}.
\bjtitle{Journal of Mechanics in Medicine and Biology}
\bvolume{15}(\bissue{5}),
\bfpage{1}--\blpage{13}
(\byear{2015})
\doiurl{10.1142/S0219519415500876}
\end{barticle}
\endbibitem

\bibitem[\protect\citeauthoryear{Doty et~al.}{2018}]{Doty_2018}
\begin{barticle}
\bauthor{\bsnm{Doty}, \binits{R.L.}},
\bauthor{\bsnm{MacGillivray}, \binits{M.R.}},
\bauthor{\bsnm{Talab}, \binits{H.}},
\bauthor{\bsnm{Tourbier}, \binits{I.}},
\bauthor{\bsnm{Reish}, \binits{M.}},
\bauthor{\bsnm{Davis}, \binits{S.}},
\bauthor{\bsnm{Cuzzocreo}, \binits{J.L.}},
\bauthor{\bsnm{Shepard}, \binits{N.T.}},
\bauthor{\bsnm{Pham}, \binits{D.L.}}:
\batitle{Balance in multiple sclerosis: relationship to central brain regions}.
\bjtitle{Experimental Brain Research}
\bvolume{236}(\bissue{10}),
\bfpage{2739}--\blpage{2750}
(\byear{2018})
\doiurl{10.1007/s00221-018-5332-1}
\end{barticle}
\endbibitem

\bibitem[\protect\citeauthoryear{Deng et~al.}{2018}]{Deng_2018}
\begin{barticle}
\bauthor{\bsnm{Deng}, \binits{W.}},
\bauthor{\bsnm{Papavasileiou}, \binits{I.}},
\bauthor{\bsnm{Qiao}, \binits{Z.}},
\bauthor{\bsnm{Zhang}, \binits{W.}},
\bauthor{\bsnm{Lam}, \binits{K.Y.}},
\bauthor{\bsnm{Han}, \binits{S.}}:
\batitle{Advances in automation technologies for lower extremity
  neurorehabilitation a review and future challenges}.
\bjtitle{IEEE Reviews in Biomedical Engineering}
\bvolume{11},
\bfpage{289}--\blpage{305}
(\byear{2018})
\doiurl{10.1109/RBME.2018.2830805}
\end{barticle}
\endbibitem

\bibitem[\protect\citeauthoryear{Hankey}{2017}]{Hankey_2017}
\begin{barticle}
\bauthor{\bsnm{Hankey}, \binits{G.J.}}:
\batitle{Stroke}.
\bjtitle{Lancet (London, England)}
\bvolume{389}(\bissue{10069}),
\bfpage{641}--\blpage{654}
(\byear{2017})
\doiurl{10.1016/S0140-6736(16)30962-X}
\end{barticle}
\endbibitem

\bibitem[\protect\citeauthoryear{Horak}{2006}]{Horak_2006}
\begin{barticle}
\bauthor{\bsnm{Horak}, \binits{F.B.}}:
\batitle{Postural orientation and equilibrium: What do we need to know about
  neural control of balance to prevent falls?}
\bjtitle{Age and Ageing}
\bvolume{35}(\bissue{SUPPL.2}),
\bfpage{7}--\blpage{11}
(\byear{2006})
\doiurl{10.1093/ageing/afl077}
\end{barticle}
\endbibitem

\bibitem[\protect\citeauthoryear{Rito et~al.}{2021}]{Rito_2021}
\begin{botherref}
\oauthor{\bsnm{Rito}, \binits{D.}},
\oauthor{\bsnm{Pinheiro}, \binits{C.}},
\oauthor{\bsnm{Figueiredo}, \binits{J.}},
\oauthor{\bsnm{Santos}, \binits{C.P.}}:
Virtual reality tools for post-stroke balance rehabilitation: A review and a
  solution proposal.
2021 IEEE International Conference on Autonomous Robot Systems and
  Competitions, ICARSC 2021,
240--245
(2021)
\doiurl{10.1109/ICARSC52212.2021.9429777}
\end{botherref}
\endbibitem

\bibitem[\protect\citeauthoryear{Aruin et~al.}{2015}]{Aruin_2015a}
\begin{botherref}
\oauthor{\bsnm{Aruin}, \binits{A.S.}},
\oauthor{\bsnm{Kanekar}, \binits{N.}},
\oauthor{\bsnm{Lee}, \binits{Y.J.}}:
Anticipatory and compensatory postural adjustments in individuals with multiple
  sclerosis in response to external perturbations.
Neuroscience Letters,
182--186
(2015)
\doiurl{10.1016/j.neulet.2015.02.050}
\end{botherref}
\endbibitem

\bibitem[\protect\citeauthoryear{Craig et~al.}{2019}]{Craig_2019}
\begin{barticle}
\bauthor{\bsnm{Craig}, \binits{J.J.}},
\bauthor{\bsnm{Bruetsch}, \binits{A.P.}},
\bauthor{\bsnm{Lynch}, \binits{S.G.}},
\bauthor{\bsnm{Huisinga}, \binits{J.M.}}:
\batitle{Altered visual and somatosensory feedback affects gait stability in
  persons with multiple sclerosis}.
\bjtitle{Human Movement Science}
\bvolume{66}(\bissue{May}),
\bfpage{355}--\blpage{362}
(\byear{2019})
\doiurl{10.1016/j.humov.2019.05.018}
\end{barticle}
\endbibitem

\bibitem[\protect\citeauthoryear{De~Angelis et~al.}{2021}]{De_2021}
\begin{botherref}
\oauthor{\bsnm{De~Angelis}, \binits{S.}},
\oauthor{\bsnm{Princi}, \binits{A.A.}},
\oauthor{\bsnm{Dal~Farra}, \binits{F.}},
\oauthor{\bsnm{Morone}, \binits{G.}},
\oauthor{\bsnm{Caltagirone}, \binits{C.}},
\oauthor{\bsnm{Tramontano}, \binits{M.}}:
Vibrotactile-based rehabilitation on balance and gait in patients with
  neurological diseases: A systematic review and metanalysis.
Brain Sciences
\textbf{11}(4)
(2021)
\doiurl{10.3390/brainsci11040518}
\end{botherref}
\endbibitem

\bibitem[\protect\citeauthoryear{Schilling et~al.}{2019}]{Schilling_2019}
\begin{barticle}
\bauthor{\bsnm{Schilling}, \binits{M.L.}},
\bauthor{\bsnm{Beattie}, \binits{L.}},
\bauthor{\bsnm{Bell}, \binits{K.}},
\bauthor{\bsnm{Jacques}, \binits{O.}},
\bauthor{\bsnm{Lyon}, \binits{C.}}:
\batitle{Effects of tai chi, fitness boxing, and video activities on the
  balance and endurance of a participant with multiple sclerosis: A case
  study}.
\bjtitle{Activities, Adaptation and Aging}
\bvolume{43}(\bissue{2}),
\bfpage{133}--\blpage{144}
(\byear{2019})
\doiurl{10.1080/01924788.2018.1500054}
\end{barticle}
\endbibitem

\bibitem[\protect\citeauthoryear{Verghese et~al.}{2010}]{PMI}
\begin{barticle}
\bauthor{\bsnm{Verghese}, \binits{J.}},
\bauthor{\bsnm{Ambrose}, \binits{A.F.}},
\bauthor{\bsnm{Lipton}, \binits{R.B.}},
\bauthor{\bsnm{Wang}, \binits{C.}}:
\batitle{Neurological gait abnormalities and risk of falls in older adults}.
\bjtitle{Journal of neurology}
\bvolume{257}(\bissue{3}),
\bfpage{392}--\blpage{398}
(\byear{2010})
\doiurl{10.1007/s00415-009-5332-y}
\end{barticle}
\endbibitem

\bibitem[\protect\citeauthoryear{Aries et~al.}{2021}]{Aries_2021}
\begin{botherref}
\oauthor{\bsnm{Aries}, \binits{A.M.}},
\oauthor{\bsnm{Pomeroy}, \binits{V.M.}},
\oauthor{\bsnm{Sim}, \binits{J.}},
\oauthor{\bsnm{Read}, \binits{S.}},
\oauthor{\bsnm{Hunter}, \binits{S.M.}}:
Sensory stimulation of the foot and ankle early post-stroke: A pilot and
  feasibility study.
Frontiers in Neurology
(July),
1--24
(2021)
\doiurl{10.3389/fneur.2021.675106}
\end{botherref}
\endbibitem

\bibitem[\protect\citeauthoryear{Kim and Jang}{2021}]{Kim_2021}
\begin{barticle}
\bauthor{\bsnm{Kim}, \binits{K.-H.}},
\bauthor{\bsnm{Jang}, \binits{S.-H.}}:
\batitle{Effects of cognitive sensory motor training on lower extremity muscle
  strength and balance in post stroke patients: A randomized controlled study}.
\bjtitle{Clinics and practice}
\bvolume{11}(\bissue{3}),
\bfpage{640}--\blpage{649}
(\byear{2021})
\doiurl{10.3390/CLINPRACT11030079}
\end{barticle}
\endbibitem

\bibitem[\protect\citeauthoryear{Costantino et~al.}{2017}]{Costantino_2017}
\begin{botherref}
\oauthor{\bsnm{Costantino}, \binits{C.}},
\oauthor{\bsnm{Galuppo}, \binits{L.}},
\oauthor{\bsnm{Romiti}, \binits{D.}}:
Short-term effect of local muscle vibration treatment versus sham therapy on
  upper limb in chronic post-stroke patients: A randomized controlled trial.
European Journal of Physical and Rehabilitation Medicine
(1),
32--40
(2017)
\doiurl{10.23736/S1973-9087.16.04211-8}
\end{botherref}
\endbibitem

\bibitem[\protect\citeauthoryear{Gorst et~al.}{2019}]{Gorst_2019}
\begin{barticle}
\bauthor{\bsnm{Gorst}, \binits{T.}},
\bauthor{\bsnm{Rogers}, \binits{A.}},
\bauthor{\bsnm{Morrison}, \binits{S.C.}},
\bauthor{\bsnm{Cramp}, \binits{M.}},
\bauthor{\bsnm{Paton}, \binits{J.}},
\bauthor{\bsnm{Freeman}, \binits{J.}},
\bauthor{\bsnm{Marsden}, \binits{J.}}:
\batitle{The prevalence, distribution, and functional importance of lower limb
  somatosensory impairments in chronic stroke survivors: a cross sectional
  observational study}.
\bjtitle{Disability and Rehabilitation}
\bvolume{41}(\bissue{20}),
\bfpage{2443}--\blpage{2450}
(\byear{2019})
\doiurl{10.1080/09638288.2018.1468932}
\end{barticle}
\endbibitem

\bibitem[\protect\citeauthoryear{Tyson et~al.}{2013}]{Tyson_2013}
\begin{barticle}
\bauthor{\bsnm{Tyson}, \binits{S.}},
\bauthor{\bsnm{Crow}, \binits{J.}},
\bauthor{\bsnm{Connell}, \binits{L.}},
\bauthor{\bsnm{Winward}, \binits{C.}},
\bauthor{\bsnm{Hillier}, \binits{S.}}:
\batitle{Sensory impairments of the lower limb after stroke: A pooled analysis
  of individual patient data}.
\bjtitle{Topics in Stroke Rehabilitation}
\bvolume{20}(\bissue{5}),
\bfpage{441}--\blpage{449}
(\byear{2013})
\doiurl{10.1310/tsr2005-441}
\end{barticle}
\endbibitem

\bibitem[\protect\citeauthoryear{Patel et~al.}{2000}]{Patel_2000}
\begin{botherref}
\oauthor{\bsnm{Patel}, \binits{A.T.}},
\oauthor{\bsnm{Duncan}, \binits{P.W.}},
\oauthor{\bsnm{Lai}, \binits{S.M.}},
\oauthor{\bsnm{Studenski}, \binits{S.}}:
The relation between impairments and functional outcomes poststroke.
Archives of Physical Medicine and Rehabilitation
(10),
1357--1363
(2000)
\doiurl{10.1053/apmr.2000.9397}
\end{botherref}
\endbibitem

\bibitem[\protect\citeauthoryear{Sánchez-Blanco et~al.}{1999}]{hez_1999}
\begin{barticle}
\bauthor{\bsnm{Sánchez-Blanco}, \binits{I.}},
\bauthor{\bsnm{Ochoa-Sangrador}, \binits{C.}},
\bauthor{\bsnm{López-Munaín}, \binits{L.}},
\bauthor{\bsnm{Izquierdo-Sánchez}, \binits{M.}},
\bauthor{\bsnm{Fermoso-García}, \binits{J.}}:
\batitle{Predictive model of functional independence in stroke patients
  admitted to a rehabilitation programme}.
\bjtitle{Clinical Rehabilitation}
\bvolume{13}(\bissue{6}),
\bfpage{464}--\blpage{475}
(\byear{1999})
\doiurl{10.1191/026921599672994947}
\end{barticle}
\endbibitem

\bibitem[\protect\citeauthoryear{Gorst et~al.}{2016}]{Gorst_2016}
\begin{barticle}
\bauthor{\bsnm{Gorst}, \binits{T.}},
\bauthor{\bsnm{Lyddon}, \binits{A.}},
\bauthor{\bsnm{Marsden}, \binits{J.}},
\bauthor{\bsnm{Paton}, \binits{J.}},
\bauthor{\bsnm{Morrison}, \binits{S.C.}},
\bauthor{\bsnm{Cramp}, \binits{M.}},
\bauthor{\bsnm{Freeman}, \binits{J.}}:
\batitle{Foot and ankle impairments affect balance and mobility in stroke
  (faimis): The views and experiences of people with stroke}.
\bjtitle{Disability and Rehabilitation}
\bvolume{38}(\bissue{6}),
\bfpage{589}--\blpage{596}
(\byear{2016})
\doiurl{10.3109/09638288.2015.1052888}
\end{barticle}
\endbibitem

\bibitem[\protect\citeauthoryear{Bowen et~al.}{2016}]{Bowen_2016}
\begin{botherref}
\oauthor{\bsnm{Bowen}, \binits{C.}},
\oauthor{\bsnm{Ashburn}, \binits{A.}},
\oauthor{\bsnm{Cole}, \binits{M.}},
\oauthor{\bsnm{Donovan-Hall}, \binits{M.}},
\oauthor{\bsnm{Burnett}, \binits{M.}},
\oauthor{\bsnm{Robison}, \binits{J.}},
\oauthor{\bsnm{Mamode}, \binits{L.}},
\oauthor{\bsnm{Pickering}, \binits{R.}},
\oauthor{\bsnm{Bader}, \binits{D.}},
\oauthor{\bsnm{Kunkel}, \binits{D.}}:
A survey exploring self-reported indoor and outdoor footwear habits, foot
  problems and fall status in people with stroke and parkinson’s.
Journal of Foot and Ankle Research
\textbf{9}(1)
(2016)
\doiurl{10.1186/s13047-016-0170-5}
\end{botherref}
\endbibitem

\bibitem[\protect\citeauthoryear{Anne Shumway-Cook}{2016}]{Anne_2016}
\begin{botherref}
\oauthor{\bsnm{Anne~Shumway-Cook}, \binits{M.H.W.}}:
Motor control: Translating research into clinical practiceno title,
680
(2016)
\end{botherref}
\endbibitem

\bibitem[\protect\citeauthoryear{Granacher et~al.}{2011}]{Granacher_2011}
\begin{barticle}
\bauthor{\bsnm{Granacher}, \binits{U.}},
\bauthor{\bsnm{Muehlbaue}, \binits{T.}},
\bauthor{\bsnm{Zahner}, \binits{L.}},
\bauthor{\bsnm{Gollhofer}, \binits{A.}},
\bauthor{\bsnm{Kressig}, \binits{R.W.}}:
\batitle{Comparison of traditional and recent approaches in the promotion of
  balance and strength in older adults}.
\bjtitle{Sports Medicine}
\bvolume{41}(\bissue{5}),
\bfpage{377}--\blpage{400}
(\byear{2011})
\doiurl{10.2165/11539920-000000000-00000}
\end{barticle}
\endbibitem

\bibitem[\protect\citeauthoryear{Lesinski et~al.}{2015}]{Lesinski_2015}
\begin{barticle}
\bauthor{\bsnm{Lesinski}, \binits{M.}},
\bauthor{\bsnm{Hortobágyi}, \binits{T.}},
\bauthor{\bsnm{Muehlbauer}, \binits{T.}},
\bauthor{\bsnm{Gollhofer}, \binits{A.}},
\bauthor{\bsnm{Granacher}, \binits{U.}}:
\batitle{Effects of balance training on balance performance in healthy older
  adults: A systematic review and meta-analysis}.
\bjtitle{Sports medicine (Auckland, N.Z.)}
\bvolume{45}(\bissue{12}),
\bfpage{1721}--\blpage{1738}
(\byear{2015})
\doiurl{10.1007/S40279-015-0375-Y}
\end{barticle}
\endbibitem

\bibitem[\protect\citeauthoryear{Aruin et~al.}{2015}]{Aruin_2015b}
\begin{botherref}
\oauthor{\bsnm{Aruin}, \binits{A.S.}},
\oauthor{\bsnm{Kanekar}, \binits{N.}},
\oauthor{\bsnm{Lee}, \binits{Y.J.}}:
Anticipatory and compensatory postural adjustments in individuals with multiple
  sclerosis in response to external perturbations.
Neuroscience Letters,
182--186
(2015)
\doiurl{10.1016/j.neulet.2015.02.050}
\end{botherref}
\endbibitem

\bibitem[\protect\citeauthoryear{Aruin et~al.}{2017}]{Aruin_2017b}
\begin{barticle}
\bauthor{\bsnm{Aruin}, \binits{A.S.}},
\bauthor{\bsnm{Ganesan}, \binits{M.}},
\bauthor{\bsnm{Lee}, \binits{Y.}}:
\batitle{Improvement of postural control in individuals with multiple sclerosis
  after a single-session of ball throwing exercise}.
\bjtitle{Multiple sclerosis and related disorders}
\bvolume{17},
\bfpage{224}--\blpage{229}
(\byear{2017})
\doiurl{10.1016/J.MSARD.2017.08.013}
\end{barticle}
\endbibitem

\bibitem[\protect\citeauthoryear{Tajali et~al.}{2018}]{Tajali_2018}
\begin{botherref}
\oauthor{\bsnm{Tajali}, \binits{S.}},
\oauthor{\bsnm{Rouhani}, \binits{M.}},
\oauthor{\bsnm{Mehravar}, \binits{M.}},
\oauthor{\bsnm{Negahban}, \binits{H.}},
\oauthor{\bsnm{Sadati}, \binits{E.}},
\oauthor{\bsnm{Oskouei}, \binits{A.E.}}:
Effects of external perturbations on anticipatory and compensatory postural
  adjustments in patients with multiple sclerosis and a fall history.
International Journal of MS Care,
164--172
(2018)
\doiurl{10.7224/1537-2073.2016-098}
\end{botherref}
\endbibitem

\bibitem[\protect\citeauthoryear{Krishnan et~al.}{2012}]{Krishnan_2012}
\begin{barticle}
\bauthor{\bsnm{Krishnan}, \binits{V.}},
\bauthor{\bsnm{Kanekar}, \binits{N.}},
\bauthor{\bsnm{Aruin}, \binits{A.S.}}:
\batitle{Anticipatory postural adjustments in individuals with multiple
  sclerosis}.
\bjtitle{Neuroscience Letters}
\bvolume{506}(\bissue{2}),
\bfpage{256}--\blpage{260}
(\byear{2012})
\doiurl{10.1016/j.neulet.2011.11.018}
\end{barticle}
\endbibitem

\bibitem[\protect\citeauthoryear{Shadmehr and Amiri}{2012}]{Shadmehr_2012}
\begin{barticle}
\bauthor{\bsnm{Shadmehr}, \binits{A.}},
\bauthor{\bsnm{Amiri}, \binits{S.}}:
\batitle{Design and construction of a computerized based systemfor reaction
  time test and anticipation skill estimation}.
\bjtitle{International Journal of Bioscience, Biochemistry and Bioinformatics}
\bvolume{2}(\bissue{6}),
\bfpage{429}--\blpage{432}
(\byear{2012})
\doiurl{10.7763/ijbbb.2012.v2.148}
\end{barticle}
\endbibitem

\bibitem[\protect\citeauthoryear{S.~Aruin}{2016}]{Aruin_2016}
\begin{barticle}
\bauthor{\bsnm{S.~Aruin}, \binits{A.}}:
\batitle{Enhancing anticipatory postural adjustments: A novel approach to
  balance rehabilitation}.
\bjtitle{Journal of Novel Physiotherapies}
\bvolume{06}(\bissue{02}),
\bfpage{10}--\blpage{12}
(\byear{2016})
\doiurl{10.4172/2165-7025.1000e144}
\end{barticle}
\endbibitem

\bibitem[\protect\citeauthoryear{Saito et~al.}{2014}]{Saito_2014}
\begin{barticle}
\bauthor{\bsnm{Saito}, \binits{H.}},
\bauthor{\bsnm{Yamanaka}, \binits{M.}},
\bauthor{\bsnm{Kasahara}, \binits{S.}},
\bauthor{\bsnm{Fukushima}, \binits{J.}}:
\batitle{Relationship between improvements in motor performance and changes in
  anticipatory postural adjustments during whole-body reaching training}.
\bjtitle{Human Movement Science}
\bvolume{37},
\bfpage{69}--\blpage{86}
(\byear{2014})
\doiurl{10.1016/j.humov.2014.07.001}
\end{barticle}
\endbibitem

\bibitem[\protect\citeauthoryear{Sandroff et~al.}{2015}]{Sandroff_2015}
\begin{barticle}
\bauthor{\bsnm{Sandroff}, \binits{B.M.}},
\bauthor{\bsnm{Hillman}, \binits{C.H.}},
\bauthor{\bsnm{Benedict}, \binits{R.H.B.}},
\bauthor{\bsnm{Motl}, \binits{R.W.}}:
\batitle{Acute effects of walking, cycling, and yoga exercise on cognition in
  persons with relapsing-remitting multiple sclerosis without impaired
  cognitive processing speed}.
\bjtitle{Journal of Clinical and Experimental Neuropsychology}
\bvolume{37}(\bissue{2}),
\bfpage{209}--\blpage{219}
(\byear{2015})
\doiurl{10.1080/13803395.2014.1001723}
\end{barticle}
\endbibitem

\bibitem[\protect\citeauthoryear{Agırbas et~al.}{2019}]{AGIRBAS_2019}
\begin{barticle}
\bauthor{\bsnm{Agırbas}, \binits{O.}},
\bauthor{\bsnm{Aggon}, \binits{E.}},
\bauthor{\bsnm{Oz}, \binits{R.}}:
\batitle{The development a new device to measure of audial and visual reaction
  time in hand and foot (validity and reliability study)}.
\bjtitle{International Education Studies}
\bvolume{12}(\bissue{4}),
\bfpage{165}
(\byear{2019})
\doiurl{10.5539/ies.v12n4p165}
\end{barticle}
\endbibitem

\bibitem[\protect\citeauthoryear{Tajali et~al.}{2019}]{Tajali_2019}
\begin{barticle}
\bauthor{\bsnm{Tajali}, \binits{S.}},
\bauthor{\bsnm{Mehravar}, \binits{M.}},
\bauthor{\bsnm{Negahban}, \binits{H.}},
\bauthor{\bsnm{Dieën}, \binits{J.H.}},
\bauthor{\bsnm{Shaterzadeh-Yazdi}, \binits{M.J.}},
\bauthor{\bsnm{Mofateh}, \binits{R.}}:
\batitle{Impaired local dynamic stability during treadmill walking predicts
  future falls in patients with multiple sclerosis: A prospective cohort
  study}.
\bjtitle{Clinical Biomechanics}
\bvolume{67}(\bissue{December 2018}),
\bfpage{197}--\blpage{201}
(\byear{2019})
\doiurl{10.1016/j.clinbiomech.2019.05.013}
\end{barticle}
\endbibitem

\bibitem[\protect\citeauthoryear{Dejanovic and
  Dejanovic}{2015}]{Dejanovic_2015}
\begin{botherref}
\oauthor{\bsnm{Dejanovic}, \binits{M.}},
\oauthor{\bsnm{Dejanovic}, \binits{I.}}:
Software for reaction-time measurement and its application for the evaluation
  of patient’s recovery after the stroke
(May 2016),
182--185
(2015)
\doiurl{10.15849/icit.2015.0027}
\end{botherref}
\endbibitem

\bibitem[\protect\citeauthoryear{Santos et~al.}{2010}]{Santos_2010}
\begin{barticle}
\bauthor{\bsnm{Santos}, \binits{M.J.}},
\bauthor{\bsnm{Kanekar}, \binits{N.}},
\bauthor{\bsnm{Aruin}, \binits{A.S.}}:
\batitle{The role of anticipatory postural adjustments in compensatory control
  of posture: 1. electromyographic analysis}.
\bjtitle{Journal of electromyography and kinesiology: official journal of the
  International Society of Electrophysiological Kinesiology}
\bvolume{20}(\bissue{3}),
\bfpage{388}--\blpage{397}
(\byear{2010})
\doiurl{10.1016/J.JELEKIN.2009.06.006}
\end{barticle}
\endbibitem

\bibitem[\protect\citeauthoryear{Aruin et~al.}{2015}]{Aruin_2015}
\begin{barticle}
\bauthor{\bsnm{Aruin}, \binits{A.S.}},
\bauthor{\bsnm{Kanekar}, \binits{N.}},
\bauthor{\bsnm{Lee}, \binits{Y.J.}},
\bauthor{\bsnm{Ganesan}, \binits{M.}}:
\batitle{Enhancement of anticipatory postural adjustments in older adults as a
  result of a single session of ball throwing exercise}.
\bjtitle{Experimental brain research}
\bvolume{233}(\bissue{2}),
\bfpage{649}--\blpage{655}
(\byear{2015})
\doiurl{10.1007/S00221-014-4144-1}
\end{barticle}
\endbibitem

\bibitem[\protect\citeauthoryear{Yamada and Shinya}{2021}]{Yamada_2021}
\begin{barticle}
\bauthor{\bsnm{Yamada}, \binits{H.}},
\bauthor{\bsnm{Shinya}, \binits{M.}}:
\batitle{Variability in the center of mass state during initiation of accurate
  forward step aimed at targets of different sizes}.
\bjtitle{Frontiers in Sports and Active Living}
\bvolume{3},
\bfpage{1}--\blpage{9}
(\byear{2021})
\doiurl{10.3389/fspor.2021.691307}
\end{barticle}
\endbibitem

\bibitem[\protect\citeauthoryear{Aruin et~al.}{2017}]{Aruin_2017a}
\begin{barticle}
\bauthor{\bsnm{Aruin}, \binits{A.S.}},
\bauthor{\bsnm{Ganesan}, \binits{M.}},
\bauthor{\bsnm{Lee}, \binits{Y.}}:
\batitle{Improvement of postural control in individuals with multiple sclerosis
  after a single-session of ball throwing exercise}.
\bjtitle{Multiple Sclerosis and Related Disorders}
\bvolume{17}(\bissue{August}),
\bfpage{224}--\blpage{229}
(\byear{2017})
\doiurl{10.1016/j.msard.2017.08.013}
\end{barticle}
\endbibitem

\bibitem[\protect\citeauthoryear{Aloraini et~al.}{2020}]{Aloraini_2020}
\begin{barticle}
\bauthor{\bsnm{Aloraini}, \binits{S.M.}},
\bauthor{\bsnm{Glazebrook}, \binits{C.M.}},
\bauthor{\bsnm{Pooyania}, \binits{S.}},
\bauthor{\bsnm{Sibley}, \binits{K.M.}},
\bauthor{\bsnm{Singer}, \binits{J.}},
\bauthor{\bsnm{Passmore}, \binits{S.}}:
\batitle{An external focus of attention compared to an internal focus of
  attention improves anticipatory postural adjustments among people
  post-stroke}.
\bjtitle{Gait and Posture}
\bvolume{82}(\bissue{July 2019}),
\bfpage{100}--\blpage{105}
(\byear{2020})
\doiurl{10.1016/j.gaitpost.2020.08.133}
\end{barticle}
\endbibitem

\bibitem[\protect\citeauthoryear{Bertucco et~al.}{2013}]{Bertucco_2013}
\begin{barticle}
\bauthor{\bsnm{Bertucco}, \binits{M.}},
\bauthor{\bsnm{Cesari}, \binits{P.}},
\bauthor{\bsnm{Latash}, \binits{M.L.}}:
\batitle{Fitts’ law in early postural adjustments}.
\bjtitle{Neuroscience}
\bvolume{231},
\bfpage{61}--\blpage{69}
(\byear{2013})
\doiurl{10.1016/j.neuroscience.2012.11.043}
\end{barticle}
\endbibitem

\bibitem[\protect\citeauthoryear{Bertucco and Cesari}{2010}]{Bertucco_2010}
\begin{barticle}
\bauthor{\bsnm{Bertucco}, \binits{M.}},
\bauthor{\bsnm{Cesari}, \binits{P.}}:
\batitle{Does movement planning follow fitts’ law? scaling anticipatory
  postural adjustments with movement speed and accuracy}.
\bjtitle{Neuroscience}
\bvolume{171}(\bissue{1}),
\bfpage{205}--\blpage{213}
(\byear{2010})
\doiurl{10.1016/j.neuroscience.2010.08.023}
\end{barticle}
\endbibitem

\bibitem[\protect\citeauthoryear{Mulder and Van~Maanen}{2013}]{Mulder_2013}
\begin{botherref}
\oauthor{\bsnm{Mulder}, \binits{M.J.}},
\oauthor{\bsnm{Van~Maanen}, \binits{L.}}:
Are accuracy and reaction time affected via different processes?
PLoS ONE
\textbf{8}(11)
(2013)
\doiurl{10.1371/journal.pone.0080222}
\end{botherref}
\endbibitem

\bibitem[\protect\citeauthoryear{Barr et~al.}{2014}]{Barr_2014}
\begin{barticle}
\bauthor{\bsnm{Barr}, \binits{C.}},
\bauthor{\bsnm{McLoughlin}, \binits{J.}},
\bauthor{\bsnm{Lord}, \binits{S.R.}},
\bauthor{\bsnm{Crotty}, \binits{M.}},
\bauthor{\bsnm{Sturnieks}, \binits{D.L.}}:
\batitle{Walking for six minutes increases both simple reaction time and
  stepping reaction time in moderately disabled people with multiple
  sclerosis}.
\bjtitle{Multiple Sclerosis and Related Disorders}
\bvolume{3}(\bissue{4}),
\bfpage{457}--\blpage{462}
(\byear{2014})
\doiurl{10.1016/j.msard.2014.01.002}
\end{barticle}
\endbibitem

\bibitem[\protect\citeauthoryear{Chien et~al.}{2017}]{Chien_2017}
\begin{barticle}
\bauthor{\bsnm{Chien}, \binits{J.H.}},
\bauthor{\bsnm{Ambati}, \binits{V.N.P.}},
\bauthor{\bsnm{Huang}, \binits{C.K.}},
\bauthor{\bsnm{Mukherjee}, \binits{M.}}:
\batitle{Tactile stimuli affect long-range correlations of stride interval and
  stride length differently during walking}.
\bjtitle{Experimental Brain Research}
\bvolume{235}(\bissue{4}),
\bfpage{1185}--\blpage{1193}
(\byear{2017})
\doiurl{10.1007/s00221-017-4881-z}
\end{barticle}
\endbibitem

\bibitem[\protect\citeauthoryear{Mansfield et~al.}{2015}]{Mansfield_2015}
\begin{barticle}
\bauthor{\bsnm{Mansfield}, \binits{A.}},
\bauthor{\bsnm{Aqui}, \binits{A.}},
\bauthor{\bsnm{Centen}, \binits{A.}},
\bauthor{\bsnm{Danells}, \binits{C.J.}},
\bauthor{\bsnm{DePaul}, \binits{V.G.}},
\bauthor{\bsnm{Knorr}, \binits{S.}},
\bauthor{\bsnm{Schinkel-Ivy}, \binits{A.}},
\bauthor{\bsnm{Brooks}, \binits{D.}},
\bauthor{\bsnm{Inness}, \binits{E.L.}},
\bauthor{\bsnm{McIlroy}, \binits{W.E.}},
\bauthor{\bsnm{al.}}:
\batitle{Perturbation training to promote safe independent mobility
  post-stroke: Study protocol for a randomized controlled trial}.
\bjtitle{BMC Neurology}
\bvolume{15}(\bissue{1}),
\bfpage{1}--\blpage{10}
(\byear{2015})
\doiurl{10.1186/S12883-015-0347-8/TABLES/2}
\end{barticle}
\endbibitem

\bibitem[\protect\citeauthoryear{Allin et~al.}{2020}]{Allin_2020}
\begin{botherref}
\oauthor{\bsnm{Allin}, \binits{L.J.}},
\oauthor{\bsnm{Brolinson}, \binits{P.G.}},
\oauthor{\bsnm{Beach}, \binits{B.M.}},
\oauthor{\bsnm{Kim}, \binits{S.}},
\oauthor{\bsnm{Nussbaum}, \binits{M.A.}},
\oauthor{\bsnm{Roberto}, \binits{K.A.}},
\oauthor{\bsnm{Madigan}, \binits{M.L.}}:
Perturbation-based balance training targeting both slip- and trip-induced falls
  among older adults: a randomized controlled trial.
BMC Geriatrics
\textbf{20}(1)
(2020)
\doiurl{10.1186/S12877-020-01605-9}
\end{botherref}
\endbibitem

\bibitem[\protect\citeauthoryear{Barzideh et~al.}{2020}]{Barzideh_2020}
\begin{barticle}
\bauthor{\bsnm{Barzideh}, \binits{A.}},
\bauthor{\bsnm{Marzolini}, \binits{S.}},
\bauthor{\bsnm{Danells}, \binits{C.}},
\bauthor{\bsnm{Jagroop}, \binits{D.}},
\bauthor{\bsnm{Huntley}, \binits{A.H.}},
\bauthor{\bsnm{Inness}, \binits{E.L.}},
\bauthor{\bsnm{Mathur}, \binits{S.}},
\bauthor{\bsnm{Mochizuki}, \binits{G.}},
\bauthor{\bsnm{Oh}, \binits{P.}},
\bauthor{\bsnm{Mansfield}, \binits{A.}}:
\batitle{Effect of reactive balance training on physical fitness poststroke:
  study protocol for a randomised non-inferiority trial}.
\bjtitle{BMJ Open}
\bvolume{10}(\bissue{6}),
\bfpage{035740}
(\byear{2020})
\doiurl{10.1136/BMJOPEN-2019-035740}
\end{barticle}
\endbibitem

\bibitem[\protect\citeauthoryear{Pai et~al.}{2014}]{Pai_2014}
\begin{barticle}
\bauthor{\bsnm{Pai}, \binits{Y.C.}},
\bauthor{\bsnm{Yang}, \binits{F.}},
\bauthor{\bsnm{Bhatt}, \binits{T.}},
\bauthor{\bsnm{Wang}, \binits{E.}}:
\batitle{Learning from laboratory-induced falling: long-term motor retention
  among older adults}.
\bjtitle{Age (Dordrecht, Netherlands)}
\bvolume{36}(\bissue{3}),
\bfpage{1367}--\blpage{1376}
(\byear{2014})
\doiurl{10.1007/S11357-014-9640-5}
\end{barticle}
\endbibitem

\bibitem[\protect\citeauthoryear{Bhatt and Pai}{2008}]{Bhatt_Pai_2008}
\begin{barticle}
\bauthor{\bsnm{Bhatt}, \binits{T.}},
\bauthor{\bsnm{Pai}, \binits{Y.C.}}:
\batitle{Immediate and latent interlimb transfer of gait stability adaptation
  following repeated exposure to slips}.
\bjtitle{Journal of motor behavior}
\bvolume{40}(\bissue{5}),
\bfpage{380}--\blpage{390}
(\byear{2008})
\doiurl{10.3200/JMBR.40.5.380-390}
\end{barticle}
\endbibitem

\bibitem[\protect\citeauthoryear{Bhatt and Pai}{2009}]{Bhatt_Pai_2009}
\begin{barticle}
\bauthor{\bsnm{Bhatt}, \binits{T.}},
\bauthor{\bsnm{Pai}, \binits{Y.C.}}:
\batitle{Generalization of gait adaptation for fall prevention: from moveable
  platform to slippery floor}.
\bjtitle{Journal of neurophysiology}
\bvolume{101}(\bissue{2}),
\bfpage{948}--\blpage{957}
(\byear{2009})
\doiurl{10.1152/JN.91004.2008}
\end{barticle}
\endbibitem

\bibitem[\protect\citeauthoryear{Gerards et~al.}{2017}]{Gerards_2017}
\begin{botherref}
\oauthor{\bsnm{Gerards}, \binits{M.H.G.}},
\oauthor{\bsnm{McCrum}, \binits{C.}},
\oauthor{\bsnm{Mansfield}, \binits{A.}},
\oauthor{\bsnm{Meijer}, \binits{K.}}:
Perturbation-based balance training for falls reduction among older adults:
  Current evidence and implications for clinical practice.
Geriatrics gerontology international
(12),
2294--2303
(2017)
\doiurl{10.1111/GGI.13082}
\end{botherref}
\endbibitem

\bibitem[\protect\citeauthoryear{Pai and Bhatt}{2007}]{Pai_2007}
\begin{barticle}
\bauthor{\bsnm{Pai}, \binits{Y.C.}},
\bauthor{\bsnm{Bhatt}, \binits{T.S.}}:
\batitle{Repeated-slip training: An emerging paradigm for prevention of
  slip-related falls among older adults}.
\bjtitle{Physical therapy}
\bvolume{87}(\bissue{11}),
\bfpage{1478}
(\byear{2007})
\doiurl{10.2522/PTJ.20060326}
\end{barticle}
\endbibitem

\bibitem[\protect\citeauthoryear{Mansfield et~al.}{2011}]{Mansfield_2011}
\begin{barticle}
\bauthor{\bsnm{Mansfield}, \binits{A.}},
\bauthor{\bsnm{Inness}, \binits{E.L.}},
\bauthor{\bsnm{Komar}, \binits{J.}},
\bauthor{\bsnm{Biasin}, \binits{L.}},
\bauthor{\bsnm{Brunton}, \binits{K.}},
\bauthor{\bsnm{Lakhani}, \binits{B.}},
\bauthor{\bsnm{Mcilroy}, \binits{W.E.}}:
\batitle{Training rapid stepping responses in an individual with stroke}.
\bjtitle{Physical therapy}
\bvolume{91}(\bissue{6}),
\bfpage{958}--\blpage{969}
(\byear{2011})
\doiurl{10.2522/PTJ.20100212}
\end{barticle}
\endbibitem

\bibitem[\protect\citeauthoryear{Aviles et~al.}{2020}]{Aviles_2020}
\begin{barticle}
\bauthor{\bsnm{Aviles}, \binits{J.}},
\bauthor{\bsnm{Wright}, \binits{D.L.}},
\bauthor{\bsnm{Allin}, \binits{L.J.}},
\bauthor{\bsnm{Alexander}, \binits{N.B.}},
\bauthor{\bsnm{Madigan}, \binits{M.L.}}:
\batitle{Improvement in trunk kinematics after treadmill-based reactive balance
  training among older adults is strongly associated with trunk kinematics
  before training}.
\bjtitle{Journal of Biomechanics}
\bvolume{113},
\bfpage{110112}
(\byear{2020})
\doiurl{10.1016/j.jbiomech.2020.110112}
\end{barticle}
\endbibitem

\bibitem[\protect\citeauthoryear{Tanvi et~al.}{2012}]{Tanvi_2012}
\begin{barticle}
\bauthor{\bsnm{Tanvi}, \binits{B.}},
\bauthor{\bsnm{Feng}, \binits{Y.}},
\bauthor{\bsnm{Yi-Chung}, \binits{P.}}:
\batitle{Learning to resist gait-slip falls: long-term retention in
  community-dwelling older adults}.
\bjtitle{Archives of physical medicine and rehabilitation}
\bvolume{93},
\bfpage{557}--\blpage{564}
(\byear{2012})
\doiurl{10.1016/J.APMR.2011.10.027}
\end{barticle}
\endbibitem

\bibitem[\protect\citeauthoryear{Gandolfi et~al.}{2015}]{Gandolfi_2015}
\begin{botherref}
\oauthor{\bsnm{Gandolfi}, \binits{M.}},
\oauthor{\bsnm{Munari}, \binits{D.}},
\oauthor{\bsnm{Geroin}, \binits{C.}},
\oauthor{\bsnm{Gajofatto}, \binits{A.}},
\oauthor{\bsnm{Benedetti}, \binits{M.D.}},
\oauthor{\bsnm{Midiri}, \binits{A.}},
\oauthor{\bsnm{Carla}, \binits{F.}},
\oauthor{\bsnm{Picelli}, \binits{A.}},
\oauthor{\bsnm{Waldner}, \binits{A.}},
\oauthor{\bsnm{Smania}, \binits{N.}}:
Sensory integration balance training in patients with multiple sclerosis: A
  randomized, controlled trial.
Multiple Sclerosis,
1453--1462
(2015)
\doiurl{10.1177/1352458514562438}
\end{botherref}
\endbibitem

\bibitem[\protect\citeauthoryear{Morone et~al.}{2019}]{Morone_2019}
\begin{barticle}
\bauthor{\bsnm{Morone}, \binits{G.}},
\bauthor{\bsnm{Spitoni}, \binits{G.F.}},
\bauthor{\bsnm{De~Bartolo}, \binits{D.}},
\bauthor{\bsnm{Ghanbari~Ghooshchy}, \binits{S.}},
\bauthor{\bsnm{Di~Iulio}, \binits{F.}},
\bauthor{\bsnm{Paolucci}, \binits{S.}},
\bauthor{\bsnm{Zoccolotti}, \binits{P.}},
\bauthor{\bsnm{Iosa}, \binits{M.}}:
\batitle{Rehabilitative devices for a top-down approach}.
\bjtitle{Expert Review of Medical Devices}
\bvolume{16}(\bissue{3}),
\bfpage{187}--\blpage{195}
(\byear{2019})
\doiurl{10.1080/17434440.2019.1574567}
\end{barticle}
\endbibitem

\bibitem[\protect\citeauthoryear{Verna et~al.}{2020}]{Verna_2020}
\begin{barticle}
\bauthor{\bsnm{Verna}, \binits{V.}},
\bauthor{\bsnm{De~Bartolo}, \binits{D.}},
\bauthor{\bsnm{Iosa}, \binits{M.}},
\bauthor{\bsnm{Fadda}, \binits{L.}},
\bauthor{\bsnm{Pinto}, \binits{G.}},
\bauthor{\bsnm{Caltagirone}, \binits{C.}},
\bauthor{\bsnm{De~Angelis}, \binits{S.}},
\bauthor{\bsnm{Tramontano}, \binits{M.}}:
\batitle{Te.m.p.o., an app for using temporal musical mismatch in post-stroke
  neurorehabilitation: A preliminary randomized controlled study}.
\bjtitle{NeuroRehabilitation}
\bvolume{47}(\bissue{2}),
\bfpage{201}--\blpage{208}
(\byear{2020})
\doiurl{10.3233/NRE-203126}
\end{barticle}
\endbibitem

\bibitem[\protect\citeauthoryear{Kearney et~al.}{2019}]{Kearney_2019}
\begin{barticle}
\bauthor{\bsnm{Kearney}, \binits{E.}},
\bauthor{\bsnm{Shellikeri}, \binits{S.}},
\bauthor{\bsnm{Martino}, \binits{R.}},
\bauthor{\bsnm{Yunusova}, \binits{Y.}}:
\batitle{Augmented visual feedback-aided interventions for motor rehabilitation
  in parkinson’s disease: a systematic review}.
\bjtitle{Disability and Rehabilitation}
\bvolume{41}(\bissue{9}),
\bfpage{995}--\blpage{1011}
(\byear{2019})
\doiurl{10.1080/09638288.2017.1419292}
\end{barticle}
\endbibitem

\bibitem[\protect\citeauthoryear{Smania et~al.}{2008}]{Smania_2008}
\begin{barticle}
\bauthor{\bsnm{Smania}, \binits{N.}},
\bauthor{\bsnm{Picelli}, \binits{A.}},
\bauthor{\bsnm{Gandolfi}, \binits{M.}},
\bauthor{\bsnm{Fiaschi}, \binits{A.}},
\bauthor{\bsnm{Tinazzi}, \binits{M.}}:
\batitle{Rehabilitation of sensorimotor integration deficits in balance
  impairment of patients with stroke hemiparesis: A before/after pilot study}.
\bjtitle{Neurological Sciences}
\bvolume{29}(\bissue{5}),
\bfpage{313}--\blpage{319}
(\byear{2008})
\doiurl{10.1007/s10072-008-0988-0}
\end{barticle}
\endbibitem

\bibitem[\protect\citeauthoryear{Derakhshanfar
  et~al.}{2021}]{Derakhshanfar_2021}
\begin{botherref}
\oauthor{\bsnm{Derakhshanfar}, \binits{M.}},
\oauthor{\bsnm{Raji}, \binits{P.}},
\oauthor{\bsnm{Bagheri}, \binits{H.}},
\oauthor{\bsnm{Jalili}, \binits{M.}},
\oauthor{\bsnm{Tarhsaz}, \binits{H.}}:
Sensory interventions on motor function, activities of daily living, and
  spasticity of the upper limb in people with stroke: A randomized clinical
  trial.
Journal of Hand Therapy
(4),
515--520
(2021)
\doiurl{10.1016/j.jht.2020.03.028}
\end{botherref}
\endbibitem

\bibitem[\protect\citeauthoryear{Kiper et~al.}{2015}]{Kiper_2015}
\begin{barticle}
\bauthor{\bsnm{Kiper}, \binits{P.}},
\bauthor{\bsnm{Baba}, \binits{A.}},
\bauthor{\bsnm{Agostini}, \binits{M.}},
\bauthor{\bsnm{Turolla}, \binits{A.}}:
\batitle{Proprioceptive based training for stroke recovery. proposal of new
  treatment modality for rehabilitation of upper limb in neurological
  diseases}.
\bjtitle{Archives of Physiotherapy}
\bvolume{5}(\bissue{1}),
\bfpage{1}--\blpage{6}
(\byear{2015})
\doiurl{10.1186/s40945-015-0007-8}
\end{barticle}
\endbibitem

\bibitem[\protect\citeauthoryear{Lim}{2019}]{Lim_2019}
\begin{barticle}
\bauthor{\bsnm{Lim}, \binits{C.}}:
\batitle{Multi-sensorimotor training improves proprioception and balance in
  subacute stroke patients: A randomized controlled pilot trial}.
\bjtitle{Frontiers in Neurology}
\bvolume{10}(\bissue{March}),
\bfpage{1}--\blpage{9}
(\byear{2019})
\doiurl{10.3389/fneur.2019.00157}
\end{barticle}
\endbibitem

\bibitem[\protect\citeauthoryear{Sigrist et~al.}{2013}]{Sigrist_2013}
\begin{barticle}
\bauthor{\bsnm{Sigrist}, \binits{R.}},
\bauthor{\bsnm{Rauter}, \binits{G.}},
\bauthor{\bsnm{Riener}, \binits{R.}},
\bauthor{\bsnm{Wolf}, \binits{P.}}:
\batitle{Augmented visual, auditory, haptic, and multimodal feedback in motor
  learning: A review}.
\bjtitle{Psychonomic Bulletin and Review}
\bvolume{20}(\bissue{1}),
\bfpage{21}--\blpage{53}
(\byear{2013})
\doiurl{10.3758/s13423-012-0333-8}
\end{barticle}
\endbibitem

\bibitem[\protect\citeauthoryear{Pan et~al.}{2019}]{Pan_2019}
\begin{barticle}
\bauthor{\bsnm{Pan}, \binits{L.}},
\bauthor{\bsnm{Zhao}, \binits{L.}},
\bauthor{\bsnm{Song}, \binits{A.}},
\bauthor{\bsnm{Yin}, \binits{Z.}},
\bauthor{\bsnm{She}, \binits{S.}}:
\batitle{A novel robot-aided upper limb rehabilitation training system based on
  multimodal feedback}.
\bjtitle{Frontiers in Robotics and AI}
\bvolume{6},
\bfpage{1}--\blpage{12}
(\byear{2019})
\doiurl{10.3389/frobt.2019.00102}
\end{barticle}
\endbibitem

\bibitem[\protect\citeauthoryear{Morone et~al.}{2021}]{Morone_2021}
\begin{barticle}
\bauthor{\bsnm{Morone}, \binits{G.}},
\bauthor{\bsnm{Ghanbari~Ghooshchy}, \binits{S.}},
\bauthor{\bsnm{Palomba}, \binits{A.}},
\bauthor{\bsnm{Baricich}, \binits{A.}},
\bauthor{\bsnm{Santamato}, \binits{A.}},
\bauthor{\bsnm{Ciritella}, \binits{C.}},
\bauthor{\bsnm{Ciancarelli}, \binits{I.}},
\bauthor{\bsnm{Molteni}, \binits{F.}},
\bauthor{\bsnm{Gimigliano}, \binits{F.}},
\bauthor{\bsnm{Iolascon}, \binits{G.}},
\bauthor{\bsnm{al.}}:
\batitle{Differentiation among bio- and augmented- feedback in technologically
  assisted rehabilitation}.
\bjtitle{Expert Review of Medical Devices}
\bvolume{18}(\bissue{6}),
\bfpage{513}--\blpage{522}
(\byear{2021})
\doiurl{10.1080/17434440.2021.1927704}
\end{barticle}
\endbibitem

\bibitem[\protect\citeauthoryear{Massetti et~al.}{2016}]{Massetti_2016}
\begin{barticle}
\bauthor{\bsnm{Massetti}, \binits{T.}},
\bauthor{\bsnm{Trevizan}, \binits{I.L.}},
\bauthor{\bsnm{Arab}, \binits{C.}},
\bauthor{\bsnm{Favero}, \binits{F.M.}},
\bauthor{\bsnm{Ribeiro-Papa}, \binits{D.C.}},
\bauthor{\bsnm{De~Mello~Monteiro}, \binits{C.B.}}:
\batitle{Virtual reality in multiple sclerosis - a systematic review}.
\bjtitle{Multiple Sclerosis and Related Disorders}
\bvolume{8},
\bfpage{107}--\blpage{112}
(\byear{2016})
\doiurl{10.1016/j.msard.2016.05.014}
\end{barticle}
\endbibitem

\bibitem[\protect\citeauthoryear{Gonçalves et~al.}{2014}]{Gon_2014}
\begin{barticle}
\bauthor{\bsnm{Gonçalves}, \binits{A.C.B.F.}},
\bauthor{\bsnm{Dos~Santos}, \binits{W.M.}},
\bauthor{\bsnm{Consoni}, \binits{L.J.}},
\bauthor{\bsnm{Siqueira}, \binits{A.A.G.}}:
\batitle{Serious games for assessment and rehabilitation of ankle movements}.
\bjtitle{SeGAH 2014 - IEEE 3rd International Conference on Serious Games and
  Applications for Health, Books of Proceedings}
(\byear{2014})
\doiurl{10.1109/SeGAH.2014.7067071}
\end{barticle}
\endbibitem

\bibitem[\protect\citeauthoryear{Maggio et~al.}{2019}]{Maggio_2019}
\begin{barticle}
\bauthor{\bsnm{Maggio}, \binits{M.G.}},
\bauthor{\bsnm{Russo}, \binits{M.}},
\bauthor{\bsnm{Cuzzola}, \binits{M.F.}},
\bauthor{\bsnm{Destro}, \binits{M.}},
\bauthor{\bsnm{La~Rosa}, \binits{G.}},
\bauthor{\bsnm{Molonia}, \binits{F.}},
\bauthor{\bsnm{Bramanti}, \binits{P.}},
\bauthor{\bsnm{Lombardo}, \binits{G.}},
\bauthor{\bsnm{De~Luca}, \binits{R.}},
\bauthor{\bsnm{Calabrò}, \binits{R.S.}}:
\batitle{Virtual reality in multiple sclerosis rehabilitation: A review on
  cognitive and motor outcomes}.
\bjtitle{Journal of Clinical Neuroscience}
\bvolume{65},
\bfpage{106}--\blpage{111}
(\byear{2019})
\doiurl{10.1016/j.jocn.2019.03.017}
\end{barticle}
\endbibitem

\bibitem[\protect\citeauthoryear{Feys and Straudi}{2019}]{Feys_2019}
\begin{barticle}
\bauthor{\bsnm{Feys}, \binits{P.}},
\bauthor{\bsnm{Straudi}, \binits{S.}}:
\batitle{Beyond therapists: Technology-aided physical ms rehabilitation
  delivery}.
\bjtitle{Multiple Sclerosis Journal}
\bvolume{25}(\bissue{10}),
\bfpage{1387}--\blpage{1393}
(\byear{2019})
\doiurl{10.1177/1352458519848968}
\end{barticle}
\endbibitem

\bibitem[\protect\citeauthoryear{Van~Breda et~al.}{2017}]{Van_2017}
\begin{botherref}
\oauthor{\bsnm{Van~Breda}, \binits{E.}},
\oauthor{\bsnm{Verwulgen}, \binits{S.}},
\oauthor{\bsnm{Saeys}, \binits{W.}},
\oauthor{\bsnm{Wuyts}, \binits{K.}},
\oauthor{\bsnm{Peeters}, \binits{T.}},
\oauthor{\bsnm{Truijen}, \binits{S.}}:
Vibrotactile feedback as a tool to improve motor learning and sports
  performance: a systematic review.
BMJ open sport exercise medicine
\textbf{3}(1)
(2017)
\doiurl{10.1136/BMJSEM-2016-000216}
\end{botherref}
\endbibitem

\bibitem[\protect\citeauthoryear{Scotto~di Luzio et~al.}{2020}]{Scotto_2020}
\begin{botherref}
\oauthor{\bsnm{Luzio}, \binits{F.}},
\oauthor{\bsnm{Lauretti}, \binits{C.}},
\oauthor{\bsnm{Cordella}, \binits{F.}},
\oauthor{\bsnm{Draicchio}, \binits{F.}},
\oauthor{\bsnm{Zollo}, \binits{L.}}:
Visual vs vibrotactile feedback for posture assessment during upper-limb
  robot-aided rehabilitation.
Applied ergonomics
\textbf{82}
(2020)
\doiurl{10.1016/J.APERGO.2019.102950}
\end{botherref}
\endbibitem

\bibitem[\protect\citeauthoryear{Hocaoglu}{2019}]{Hocaoglu_2019}
\begin{barticle}
\bauthor{\bsnm{Hocaoglu}, \binits{E.}}:
\batitle{Wefits: Wearable fingertip tactile sensor}.
\bjtitle{Biosystems and Biorobotics}
\bvolume{22},
\bfpage{28}--\blpage{32}
(\byear{2019})
\doiurl{10.1007/978-3-030-01887-0_6/COVER}
\end{barticle}
\endbibitem

\bibitem[\protect\citeauthoryear{Afzal et~al.}{2018}]{Afzal_2018}
\begin{barticle}
\bauthor{\bsnm{Afzal}, \binits{M.R.}},
\bauthor{\bsnm{Pyo}, \binits{S.}},
\bauthor{\bsnm{Oh}, \binits{M.K.}},
\bauthor{\bsnm{Park}, \binits{Y.S.}},
\bauthor{\bsnm{Yoon}, \binits{J.}}:
\batitle{Evaluating the effects of delivering integrated kinesthetic and
  tactile cues to individuals with unilateral hemiparetic stroke during
  overground walking}.
\bjtitle{Journal of NeuroEngineering and Rehabilitation}
\bvolume{15}(\bissue{1}),
\bfpage{1}--\blpage{14}
(\byear{2018})
\doiurl{10.1186/s12984-018-0372-0}
\end{barticle}
\endbibitem

\bibitem[\protect\citeauthoryear{Lee et~al.}{2015}]{Lee_2015}
\begin{barticle}
\bauthor{\bsnm{Lee}, \binits{B.C.}},
\bauthor{\bsnm{Thrasher}, \binits{T.A.}},
\bauthor{\bsnm{Fisher}, \binits{S.P.}},
\bauthor{\bsnm{Layne}, \binits{C.S.}}:
\batitle{The effects of different sensory augmentation on weight-shifting
  balance exercises in parkinson’s disease and healthy elderly people: A
  proof-of-concept study}.
\bjtitle{Journal of NeuroEngineering and Rehabilitation}
(\bissue{1})
(\byear{2015})
\doiurl{10.1186/s12984-015-0064-y}
\end{barticle}
\endbibitem

\bibitem[\protect\citeauthoryear{Otis et~al.}{2016}]{Otis_2016}
\begin{botherref}
\oauthor{\bsnm{Otis}, \binits{M.J.D.}},
\oauthor{\bsnm{Ayena}, \binits{J.C.}},
\oauthor{\bsnm{Tremblay}, \binits{L.E.}},
\oauthor{\bsnm{Fortin}, \binits{P.E.}},
\oauthor{\bsnm{Ménélas}, \binits{B.A.J.}}:
Use of an enactive insole for reducing the risk of falling on different types
  of soil using vibrotactile cueing for the elderly.
PLoS ONE
\textbf{11}(9)
(2016)
\doiurl{10.1371/journal.pone.0162107}
\end{botherref}
\endbibitem

\bibitem[\protect\citeauthoryear{Carey et~al.}{2016}]{Carey_2016}
\begin{botherref}
\oauthor{\bsnm{Carey}, \binits{L.M.}},
\oauthor{\bsnm{Abbott}, \binits{D.F.}},
\oauthor{\bsnm{Lamp}, \binits{G.}},
\oauthor{\bsnm{Puce}, \binits{A.}},
\oauthor{\bsnm{Seitz}, \binits{R.J.}},
\oauthor{\bsnm{Donnan}, \binits{G.A.}}:
Same intervention-different reorganization: The impact of lesion location on
  training-facilitated somatosensory recovery after stroke.
Neurorehabilitation and Neural Repair,
988--1000
(2016)
\doiurl{10.1177/1545968316653836}
\end{botherref}
\endbibitem

\bibitem[\protect\citeauthoryear{Meyer et~al.}{2016}]{Meyer_2016}
\begin{botherref}
\oauthor{\bsnm{Meyer}, \binits{S.}},
\oauthor{\bsnm{De~Bruyn}, \binits{N.}},
\oauthor{\bsnm{Lafosse}, \binits{C.}},
\oauthor{\bsnm{Van~Dijk}, \binits{M.}},
\oauthor{\bsnm{Michielsen}, \binits{M.}},
\oauthor{\bsnm{Thijs}, \binits{L.}},
\oauthor{\bsnm{Truyens}, \binits{V.}},
\oauthor{\bsnm{Oostra}, \binits{K.}},
\oauthor{\bsnm{Krumlinde-Sundholm}, \binits{L.}},
\oauthor{\bsnm{Peeters}, \binits{A.}},
\oauthor{\bsnm{al.}}:
Somatosensory impairments in the upper limb poststroke: Distribution and
  association with motor function and visuospatial neglect.
Neurorehabilitation and Neural Repair,
731--742
(2016)
\doiurl{10.1177/1545968315624779}
\end{botherref}
\endbibitem

\bibitem[\protect\citeauthoryear{Shakti et~al.}{2018}]{Shakti_2018}
\begin{botherref}
\oauthor{\bsnm{Shakti}, \binits{D.}},
\oauthor{\bsnm{Mathew}, \binits{L.}},
\oauthor{\bsnm{Kumar}, \binits{N.}},
\oauthor{\bsnm{Kataria}, \binits{C.}}:
Effectiveness of robo-assisted lower limb rehabilitation for spastic patients:
  A systematic review.
Biosensors and Bioelectronics
(April),
403--415
(2018)
\doiurl{10.1016/j.bios.2018.06.027}
\end{botherref}
\endbibitem

\bibitem[\protect\citeauthoryear{Saglia et~al.}{2009}]{Saglia_2009}
\begin{botherref}
\oauthor{\bsnm{Saglia}, \binits{J.A.}},
\oauthor{\bsnm{Tsagarakis}, \binits{N.G.}},
\oauthor{\bsnm{Dai}, \binits{J.S.}},
\oauthor{\bsnm{Caldwell}, \binits{D.G.}}:
A high performance 2-dof over-actuated parallel mechanism for ankle
  rehabilitation.
Proceedings - IEEE International Conference on Robotics and Automation,
2180--2186
(2009)
\doiurl{10.1109/ROBOT.2009.5152604}
\end{botherref}
\endbibitem

\bibitem[\protect\citeauthoryear{Saglia et~al.}{2010}]{Saglia_2010}
\begin{botherref}
\oauthor{\bsnm{Saglia}, \binits{J.A.}},
\oauthor{\bsnm{Tsagarakis}, \binits{N.G.}},
\oauthor{\bsnm{Dai}, \binits{J.S.}},
\oauthor{\bsnm{Caldwell}, \binits{D.G.}}:
Control strategies for ankle rehabilitation using a high performance ankle
  exerciser.
Proceedings - IEEE International Conference on Robotics and Automation,
2221--2227
(2010)
\doiurl{10.1109/ROBOT.2010.5509883}
\end{botherref}
\endbibitem

\bibitem[\protect\citeauthoryear{Krebs and Volpe}{2013}]{Krebs_2013}
\begin{bbook}
\bauthor{\bsnm{Krebs}, \binits{H.I.}},
\bauthor{\bsnm{Volpe}, \binits{B.T.}}:
\bbtitle{Rehabilitation Robotics}
vol. \bseriesno{110},
\bedition{1}st edn.
\bpublisher{Elsevier B.V.}, \blocation{???}
(\byear{2013}).
\doiurl{10.1016/B978-0-444-52901-5.00023-X}
\end{bbook}
\endbibitem

\bibitem[\protect\citeauthoryear{Kalita et~al.}{2021}]{Kalita_2021}
\begin{botherref}
\oauthor{\bsnm{Kalita}, \binits{B.}},
\oauthor{\bsnm{Narayan}, \binits{J.}},
\oauthor{\bsnm{Dwivedy}, \binits{S.K.}}:
Development of active lower limb robotic-based orthosis and exoskeleton
  devices: A systematic review.
International Journal of Social Robotics,
775--793
(2021)
\doiurl{10.1007/s12369-020-00662-9}
\end{botherref}
\endbibitem

\bibitem[\protect\citeauthoryear{Díaz et~al.}{2011}]{Daz_2011}
\begin{botherref}
\oauthor{\bsnm{Díaz}, \binits{I.}},
\oauthor{\bsnm{Gil}, \binits{J.J.}},
\oauthor{\bsnm{Sánchez}, \binits{E.}}:
Lower-limb robotic rehabilitation: Literature review and challenges.
Journal of Robotics,
1--11
(2011)
\doiurl{10.1155/2011/759764}
\end{botherref}
\endbibitem

\bibitem[\protect\citeauthoryear{Yurkewich et~al.}{2015}]{Yurkewich_2015}
\begin{botherref}
\oauthor{\bsnm{Yurkewich}, \binits{A.}},
\oauthor{\bsnm{Atashzar}, \binits{S.F.}},
\oauthor{\bsnm{Ayad}, \binits{A.}},
\oauthor{\bsnm{Patel}, \binits{R.V.}}:
A six-degree-of-freedom robotic system for lower extremity rehabilitation.
IEEE International Conference on Rehabilitation Robotics,
810--815
(2015)
\doiurl{10.1109/ICORR.2015.7281302}
\end{botherref}
\endbibitem

\bibitem[\protect\citeauthoryear{Valles et~al.}{2017}]{Val_2017}
\begin{barticle}
\bauthor{\bsnm{Valles}, \binits{M.}},
\bauthor{\bsnm{Cazalilla}, \binits{J.}},
\bauthor{\bsnm{Valera}, \binits{a.}},
\bauthor{\bsnm{Mata}, \binits{V.}},
\bauthor{\bsnm{Page}, \binits{a.}},
\bauthor{\bsnm{DIaz-Rodriguez}, \binits{M.}}:
\batitle{A 3-prs parallel manipulator for ankle rehabilitation: Towards a
  low-cost robotic rehabilitation}.
\bjtitle{Robotica}
\bvolume{35},
\bfpage{1939}--\blpage{1957}
(\byear{2017})
\doiurl{10.1017/S0263574715000120}
\end{barticle}
\endbibitem

\bibitem[\protect\citeauthoryear{Bernhardt et~al.}{2005}]{Bernhardt_2005}
\begin{barticle}
\bauthor{\bsnm{Bernhardt}, \binits{M.}},
\bauthor{\bsnm{Frey}, \binits{M.}},
\bauthor{\bsnm{Colombo}, \binits{G.}},
\bauthor{\bsnm{Riener}, \binits{R.}}:
\batitle{Hybrid force-position control yields cooperative behaviour of the
  rehabilitation robot lokomat}.
\bjtitle{Proceedings of the 2005 IEEE 9th International Conference on
  Rehabilitation Robotics}
\bvolume{2005},
\bfpage{536}--\blpage{539}
(\byear{2005})
\doiurl{10.1109/ICORR.2005.1501159}
\end{barticle}
\endbibitem

\bibitem[\protect\citeauthoryear{Teramae et~al.}{2018}]{Teramae_2018}
\begin{botherref}
\oauthor{\bsnm{Teramae}, \binits{T.}},
\oauthor{\bsnm{Noda}, \binits{T.}},
\oauthor{\bsnm{Morimoto}, \binits{J.}}:
Emg-based model predictive control for physical human-robot interaction:
  Application for assist-as-needed control.
IEEE Robotics and Automation,
210--217
(2018)
\doiurl{10.1109/LRA.2017.2737478}
\end{botherref}
\endbibitem

\bibitem[\protect\citeauthoryear{Li et~al.}{2017}]{Li_2017}
\begin{botherref}
\oauthor{\bsnm{Li}, \binits{B.}},
\oauthor{\bsnm{Li}, \binits{G.}},
\oauthor{\bsnm{Sun}, \binits{Y.}},
\oauthor{\bsnm{Jiang}, \binits{G.}},
\oauthor{\bsnm{Kong}, \binits{J.}},
\oauthor{\bsnm{Jiang}, \binits{D.}}:
A review of rehabilitation robot.
Proceedings - 2017 32nd Youth Academic Annual Conference of Chinese Association
  of Automation,
907--911
(2017)
\doiurl{10.1109/YAC.2017.7967538}
\end{botherref}
\endbibitem

\bibitem[\protect\citeauthoryear{Dong et~al.}{2021}]{Dong_2021}
\begin{botherref}
\oauthor{\bsnm{Dong}, \binits{M.}},
\oauthor{\bsnm{Zhou}, \binits{Y.}},
\oauthor{\bsnm{Li}, \binits{J.}},
\oauthor{\bsnm{Rong}, \binits{X.}},
\oauthor{\bsnm{Fan}, \binits{W.}},
\oauthor{\bsnm{Zhou}, \binits{X.}},
\oauthor{\bsnm{Kong}, \binits{Y.}}:
State of the art in parallel ankle rehabilitation robot: a systematic review.
Journal of NeuroEngineering and Rehabilitation
\textbf{18}
(2021)
\doiurl{10.1186/S12984-021-00845-Z}
\end{botherref}
\endbibitem

\bibitem[\protect\citeauthoryear{Marchal-Crespo and
  Reinkensmeyer}{2009}]{Marchal_2009}
\begin{barticle}
\bauthor{\bsnm{Marchal-Crespo}, \binits{L.}},
\bauthor{\bsnm{Reinkensmeyer}, \binits{D.J.}}:
\batitle{Review of control strategies for robotic movement training after
  neurologic injury}.
\bjtitle{Journal of NeuroEngineering and Rehabilitation}
(\bissue{1})
(\byear{2009})
\doiurl{10.1186/1743-0003-6-20}
\end{barticle}
\endbibitem

\bibitem[\protect\citeauthoryear{Jin et~al.}{2018}]{Jin_2018}
\begin{barticle}
\bauthor{\bsnm{Jin}, \binits{L.}},
\bauthor{\bsnm{Li}, \binits{S.}},
\bauthor{\bsnm{Yu}, \binits{J.}},
\bauthor{\bsnm{He}, \binits{J.}}:
\batitle{Robot manipulator control using neural networks: A survey}.
\bjtitle{Neurocomputing}
\bvolume{285},
\bfpage{23}--\blpage{34}
(\byear{2018})
\doiurl{10.1016/J.NEUCOM.2018.01.002}
\end{barticle}
\endbibitem

\bibitem[\protect\citeauthoryear{Dzahir and Yamamoto}{2014}]{Dzahir_2014}
\begin{barticle}
\bauthor{\bsnm{Dzahir}, \binits{M.A.M.}},
\bauthor{\bsnm{Yamamoto}, \binits{S.I.}}:
\batitle{Recent trends in lower-limb robotic rehabilitation orthosis: Control
  scheme and strategy for pneumatic muscle actuated gait trainers}.
\bjtitle{Robotics 2014, Vol. 3, Pages 120-148}
\bvolume{3}(\bissue{2}),
\bfpage{120}--\blpage{148}
(\byear{2014})
\doiurl{10.3390/ROBOTICS3020120}
\end{barticle}
\endbibitem

\bibitem[\protect\citeauthoryear{Mohebbi}{2020}]{Mohebbi_2020}
\begin{barticle}
\bauthor{\bsnm{Mohebbi}, \binits{A.}}:
\batitle{Human-robot interaction in rehabilitation and assistance: a review}.
\bjtitle{Current Robotics Reports 2020 1:3}
\bvolume{1},
\bfpage{131}--\blpage{144}
(\byear{2020})
\doiurl{10.1007/S43154-020-00015-4}
\end{barticle}
\endbibitem

\bibitem[\protect\citeauthoryear{Ayas and Altas}{2017}]{Ayas_2017}
\begin{barticle}
\bauthor{\bsnm{Ayas}, \binits{M.S.}},
\bauthor{\bsnm{Altas}, \binits{I.H.}}:
\batitle{A redundantly actuated ankle rehabilitation robot and its control
  strategies}.
\bjtitle{2016 IEEE Symposium Series on Computational Intelligence, SSCI 2016}
(\byear{2017})
\doiurl{10.1109/SSCI.2016.7850068}
\end{barticle}
\endbibitem

\bibitem[\protect\citeauthoryear{Girone et~al.}{1999}]{girone1999rutgers}
\begin{botherref}
\oauthor{\bsnm{Girone}, \binits{M.J.}},
\oauthor{\bsnm{Burdea}, \binits{G.C.}},
\oauthor{\bsnm{Bouzit}, \binits{M.}}:
The “rutgers ankle” orthopedic rehabilitation interface,
305--312
(1999).
American Society of Mechanical Engineers
\end{botherref}
\endbibitem

\bibitem[\protect\citeauthoryear{Zhang et~al.}{2016}]{Zhang_2016}
\begin{barticle}
\bauthor{\bsnm{Zhang}, \binits{F.}},
\bauthor{\bsnm{Hou}, \binits{Z.G.}},
\bauthor{\bsnm{Cheng}, \binits{L.}},
\bauthor{\bsnm{Wang}, \binits{W.}},
\bauthor{\bsnm{Chen}, \binits{Y.}},
\bauthor{\bsnm{Hu}, \binits{J.}},
\bauthor{\bsnm{Peng}, \binits{L.}},
\bauthor{\bsnm{Wang}, \binits{H.}}:
\batitle{Ileg-a lower limb rehabilitation robot: A proof of concept}.
\bjtitle{IEEE Transactions on Human-Machine Systems}
\bvolume{46}(\bissue{5}),
\bfpage{761}--\blpage{768}
(\byear{2016})
\doiurl{10.1109/THMS.2016.2562510}
\end{barticle}
\endbibitem

\bibitem[\protect\citeauthoryear{Saglia et~al.}{2013}]{Saglia_2013}
\begin{barticle}
\bauthor{\bsnm{Saglia}, \binits{J.A.}},
\bauthor{\bsnm{Tsagarakis}, \binits{N.G.}},
\bauthor{\bsnm{Dai}, \binits{J.S.}},
\bauthor{\bsnm{Caldwell}, \binits{D.G.}}:
\batitle{Control strategies for patient-assisted training using the ankle
  rehabilitation robot (arbot)}.
\bjtitle{IEEE/ASME Transactions on Mechatronics}
\bvolume{18}(\bissue{6}),
\bfpage{1799}--\blpage{1808}
(\byear{2013})
\doiurl{10.1109/TMECH.2012.2214228}
\end{barticle}
\endbibitem

\bibitem[\protect\citeauthoryear{Lu et~al.}{2016}]{lu_2016}
\begin{barticle}
\bauthor{\bsnm{Lu}, \binits{Z.}},
\bauthor{\bsnm{Li}, \binits{W.}},
\bauthor{\bsnm{Li}, \binits{M.}},
\bauthor{\bsnm{Wu}, \binits{Z.}},
\bauthor{\bsnm{Duan}, \binits{L.}},
\bauthor{\bsnm{Li}, \binits{Z.}},
\bauthor{\bsnm{Ou}, \binits{X.}},
\bauthor{\bsnm{Wang}, \binits{C.}},
\bauthor{\bsnm{Wang}, \binits{L.}},
\bauthor{\bsnm{Qin}, \binits{J.}},
\bauthor{\bsnm{al.}}:
\batitle{Development of a three freedoms ankle rehabilitation robot for ankle
  training}.
\bjtitle{IEEE Region 10 Annual International Conference, Proceedings/TENCON}
(\byear{2016})
\doiurl{10.1109/TENCON.2015.7372915}
\end{barticle}
\endbibitem

\bibitem[\protect\citeauthoryear{Song et~al.}{2019}]{Song_2019}
\begin{barticle}
\bauthor{\bsnm{Song}, \binits{P.}},
\bauthor{\bsnm{Yu}, \binits{Y.}},
\bauthor{\bsnm{Zhang}, \binits{X.}}:
\batitle{A tutorial survey and comparison of impedance control on robotic
  manipulation}.
\bjtitle{Robotica}
\bvolume{37}(\bissue{5}),
\bfpage{801}--\blpage{836}
(\byear{2019})
\doiurl{10.1017/S0263574718001339}
\end{barticle}
\endbibitem

\bibitem[\protect\citeauthoryear{Jamwal et~al.}{2016}]{Jamwal_2016}
\begin{barticle}
\bauthor{\bsnm{Jamwal}, \binits{P.K.}},
\bauthor{\bsnm{Hussain}, \binits{S.}},
\bauthor{\bsnm{Ghayesh}, \binits{M.H.}},
\bauthor{\bsnm{Rogozina}, \binits{S.V.}}:
\batitle{Impedance control of an intrinsically compliant parallel ankle
  rehabilitation robot}.
\bjtitle{IEEE Transactions on Industrial Electronics}
\bvolume{63}(\bissue{6}),
\bfpage{3638}--\blpage{3647}
(\byear{2016})
\doiurl{10.1109/TIE.2016.2521600}
\end{barticle}
\endbibitem

\bibitem[\protect\citeauthoryear{Ibarra and Siqueira}{2015}]{Ibarra_2015}
\begin{botherref}
\oauthor{\bsnm{Ibarra}, \binits{J.C.P.}},
\oauthor{\bsnm{Siqueira}, \binits{A.A.G.}}:
Impedance control of rehabilitation robots for lower limbs. review.
Proceedings - 2nd SBR Brazilian Robotics Symposium, 11th LARS Latin American
  Robotics Symposium and 6th Robocontrol Workshop on Applied Robotics and
  Automation, SBR LARS Robocontrol 2014 - Part of the Joint Conference on
  Robotics and Intelligent Systems,,
235--240
(2015)
\doiurl{10.1109/SBR.LARS.Robocontrol.2014.53}
\end{botherref}
\endbibitem

\bibitem[\protect\citeauthoryear{Sun et~al.}{2015}]{Sun_2015}
\begin{botherref}
\oauthor{\bsnm{Sun}, \binits{T.}},
\oauthor{\bsnm{Lu}, \binits{Z.}},
\oauthor{\bsnm{Wang}, \binits{C.}},
\oauthor{\bsnm{Duan}, \binits{L.}},
\oauthor{\bsnm{Shen}, \binits{Y.}},
\oauthor{\bsnm{Shi}, \binits{Q.}},
\oauthor{\bsnm{Wei}, \binits{J.}},
\oauthor{\bsnm{Wang}, \binits{Y.}},
\oauthor{\bsnm{Li}, \binits{W.}},
\oauthor{\bsnm{Qin}, \binits{J.}},
\oauthor{\bsnm{al.}}:
Mechanism design and control strategies of an ankle robot for rehabilitation
  training.
IEEE International Conference on Robotics and Biomimetics, IEEE-ROBIO 2015,
132--137
(2015)
\doiurl{10.1109/ROBIO.2015.7418756}
\end{botherref}
\endbibitem

\bibitem[\protect\citeauthoryear{Codourey}{2016}]{Codourey_2016}
\begin{botherref}
\oauthor{\bsnm{Codourey}, \binits{A.}}:
Dynamic modeling of parallel robots for computed-torque control implementation:
\textbf{17}(12),
1325--1336
(2016)
\doiurl{10.1177/027836499801701205}
\end{botherref}
\endbibitem

\bibitem[\protect\citeauthoryear{Tsoi et~al.}{2009}]{Tsoi_2009}
\begin{barticle}
\bauthor{\bsnm{Tsoi}, \binits{Y.H.}},
\bauthor{\bsnm{Xie}, \binits{S.Q.}},
\bauthor{\bsnm{Graham}, \binits{A.E.}}:
\batitle{Design, modeling and control of an ankle rehabilitation robot}.
\bjtitle{Studies in Computational Intelligence}
\bvolume{177},
\bfpage{377}--\blpage{399}
(\byear{2009})
\doiurl{10.1007/978-3-540-89933-4_18}
\end{barticle}
\endbibitem

\bibitem[\protect\citeauthoryear{Asgari and Ardestani}{2015}]{Asgari_2015}
\begin{barticle}
\bauthor{\bsnm{Asgari}, \binits{M.}},
\bauthor{\bsnm{Ardestani}, \binits{M.A.}}:
\batitle{Dynamics and improved computed torque control of a novel medical
  parallel manipulator: Applied to chest compressions to assist in
  cardiopulmonary resuscitation}.
\bjtitle{Journal of Mechanics in Medicine and Biology}
\bvolume{15}(\bissue{4}),
\bfpage{1}--\blpage{23}
(\byear{2015})
\doiurl{10.1142/S0219519415500517}
\end{barticle}
\endbibitem

\bibitem[\protect\citeauthoryear{Gao et~al.}{2014}]{Gao_2014}
\begin{barticle}
\bauthor{\bsnm{Gao}, \binits{S.H.}},
\bauthor{\bsnm{Fan}, \binits{R.}},
\bauthor{\bsnm{Wang}, \binits{D.}}:
\batitle{Dynamic modeling and model-based force control of a 3-dof
  translational parallel robot}.
\bjtitle{Advanced Materials Research}
\bvolume{1006–1007},
\bfpage{609}--\blpage{617}
(\byear{2014})
\doiurl{10.4028/www.scientific.net/AMR.1006-1007.609}
\end{barticle}
\endbibitem

\bibitem[\protect\citeauthoryear{Wolbrecht et~al.}{2008}]{Wolbrecht_2008}
\begin{barticle}
\bauthor{\bsnm{Wolbrecht}, \binits{E.T.}},
\bauthor{\bsnm{Chan}, \binits{V.}},
\bauthor{\bsnm{Reinkensmeyer}, \binits{D.J.}},
\bauthor{\bsnm{Bobrow}, \binits{J.E.}}:
\batitle{Optimizing compliant, model-based robotic assistance to promote
  neurorehabilitation}.
\bjtitle{IEEE Transactions on Neural Systems and Rehabilitation Engineering}
\bvolume{16}(\bissue{3}),
\bfpage{286}--\blpage{297}
(\byear{2008})
\doiurl{10.1109/TNSRE.2008.918389}
\end{barticle}
\endbibitem

\bibitem[\protect\citeauthoryear{Sharma and Obaid}{2020}]{Sharma_2020}
\begin{barticle}
\bauthor{\bsnm{Sharma}, \binits{S.}},
\bauthor{\bsnm{Obaid}, \binits{A.J.}}:
\batitle{Mathematical modelling, analysis and design of fuzzy logic controller
  for the control of ventilation systems using matlab fuzzy logic toolbox}.
\bjtitle{Journal of Interdisciplinary Mathematics}
\bvolume{23}(\bissue{4}),
\bfpage{843}--\blpage{849}
(\byear{2020})
\doiurl{10.1080/09720502.2020.1727611}
\end{barticle}
\endbibitem

\bibitem[\protect\citeauthoryear{Karasakal et~al.}{2005}]{Karasakal_2005}
\begin{botherref}
\oauthor{\bsnm{Karasakal}, \binits{O.}},
\oauthor{\bsnm{Yeşil}, \binits{E.}},
\oauthor{\bsnm{Güzelkaya}, \binits{M.}},
\oauthor{\bsnm{Eksin}, \binits{I.}}:
Implementation of a new self-tuning fuzzy pid controller on plc.
Turkish Journal of Electrical Engineering and Computer Sciences,
277--286
(2005)
\end{botherref}
\endbibitem

\bibitem[\protect\citeauthoryear{Lamamra et~al.}{2020}]{Lamamra_2020}
\begin{botherref}
\oauthor{\bsnm{Lamamra}, \binits{K.}},
\oauthor{\bsnm{Batat}, \binits{F.}},
\oauthor{\bsnm{Mokhtari}, \binits{F.}}:
A new technique with improved control quality of nonlinear systems using an
  optimized fuzzy logic controller.
Expert Systems with Applications
\textbf{145}
(2020)
\doiurl{10.1016/j.eswa.2019.113148}
\end{botherref}
\endbibitem

\bibitem[\protect\citeauthoryear{Patarinski and Botev}{1993}]{Patarinski_1993}
\begin{barticle}
\bauthor{\bsnm{Patarinski}, \binits{S.P.}},
\bauthor{\bsnm{Botev}, \binits{R.G.}}:
\batitle{Robot force control: A review}.
\bjtitle{Mechatronics}
\bvolume{3}(\bissue{4}),
\bfpage{377}--\blpage{398}
(\byear{1993})
\doiurl{10.1016/0957-4158(93)90012-Q}
\end{barticle}
\endbibitem

\bibitem[\protect\citeauthoryear{Chiaverini and
  Sciavicco}{1993}]{Chiaverini_1993}
\begin{botherref}
\oauthor{\bsnm{Chiaverini}, \binits{S.}},
\oauthor{\bsnm{Sciavicco}, \binits{L.}}:
The parallel approach to force/position control of robotic manipulators
\textbf{9},
361--373
(1993)
\doiurl{10.1109/70.246048}
\end{botherref}
\endbibitem

\bibitem[\protect\citeauthoryear{Keller et~al.}{2013}]{Keller_2013}
\begin{botherref}
\oauthor{\bsnm{Keller}, \binits{U.}},
\oauthor{\bsnm{Rauter}, \binits{G.}},
\oauthor{\bsnm{Riener}, \binits{R.}}:
Assist-as-needed path control for the pascal rehabilitation robot.
IEEE International Conference on Rehabilitation Robotics,
1--7
(2013)
\doiurl{10.1109/ICORR.2013.6650475}
\end{botherref}
\endbibitem

\bibitem[\protect\citeauthoryear{Pehlivan et~al.}{2017}]{Pehlivan_2017}
\begin{botherref}
\oauthor{\bsnm{Pehlivan}, \binits{A.U.}},
\oauthor{\bsnm{Losey}, \binits{D.P.}},
\oauthor{\bsnm{Rose}, \binits{C.G.}},
\oauthor{\bsnm{O’Malley}, \binits{M.K.}}:
Maintaining subject engagement during robotic rehabilitation with a minimal
  assist-as-needed (maan) controller.
IEEE International Conference on Rehabilitation Robotics,
62--67
(2017)
\doiurl{10.1109/ICORR.2017.8009222}
\end{botherref}
\endbibitem

\bibitem[\protect\citeauthoryear{Luo et~al.}{2019}]{Luo_2019}
\begin{barticle}
\bauthor{\bsnm{Luo}, \binits{L.}},
\bauthor{\bsnm{Peng}, \binits{L.}},
\bauthor{\bsnm{Wang}, \binits{C.}},
\bauthor{\bsnm{Hou}, \binits{Z.G.}}:
\batitle{A greedy assist-as-needed controller for upper limb rehabilitation}.
\bjtitle{IEEE Transactions on Neural Networks and Learning Systems}
\bvolume{30}(\bissue{11}),
\bfpage{3433}--\blpage{3443}
(\byear{2019})
\doiurl{10.1109/TNNLS.2019.2892157}
\end{barticle}
\endbibitem

\bibitem[\protect\citeauthoryear{ERSOY and HOCAOGLU}{2023}]{Ersoy_2021}
\begin{barticle}
\bauthor{\bsnm{ERSOY}, \binits{T.}},
\bauthor{\bsnm{HOCAOGLU}, \binits{E.}}:
\batitle{A 3-dof robotic platform for the rehabilitation and assessment of
  reaction time and balance skills of ms patients}.
\bjtitle{PLOS ONE}
\bvolume{18}(\bissue{2}),
\bfpage{1}--\blpage{36}
(\byear{2023})
\doiurl{10.1371/journal.pone.0280505}
\end{barticle}
\endbibitem

\bibitem[\protect\citeauthoryear{Dollar and Herr}{2008}]{Dollar_2008}
\begin{barticle}
\bauthor{\bsnm{Dollar}, \binits{A.M.}},
\bauthor{\bsnm{Herr}, \binits{H.}}:
\batitle{Lower extremity exoskeletons and active orthoses: Challenges and
  state-of-the-art}.
\bjtitle{IEEE Transactions on Robotics}
\bvolume{24}(\bissue{1}),
\bfpage{144}--\blpage{158}
(\byear{2008})
\doiurl{10.1109/TRO.2008.915453}
\end{barticle}
\endbibitem

\bibitem[\protect\citeauthoryear{Ferris et~al.}{2006}]{Ferris_2006}
\begin{barticle}
\bauthor{\bsnm{Ferris}, \binits{D.P.}},
\bauthor{\bsnm{Gordon}, \binits{K.E.}},
\bauthor{\bsnm{Sawicki}, \binits{G.S.}},
\bauthor{\bsnm{Peethambaran}, \binits{A.}}:
\batitle{An improved powered ankle–foot orthosis using proportional
  myoelectric control}.
\bjtitle{Gait Posture}
\bvolume{23}(\bissue{4}),
\bfpage{425}--\blpage{428}
(\byear{2006})
\doiurl{10.1016/J.GAITPOST.2005.05.004}
\end{barticle}
\endbibitem

\bibitem[\protect\citeauthoryear{Mooney et~al.}{2014}]{Mooney_2014}
\begin{barticle}
\bauthor{\bsnm{Mooney}, \binits{L.M.}},
\bauthor{\bsnm{Rouse}, \binits{E.J.}},
\bauthor{\bsnm{Herr}, \binits{H.M.}}:
\batitle{Autonomous exoskeleton reduces metabolic cost of human walking during
  load carriage}.
\bjtitle{Journal of NeuroEngineering and Rehabilitation}
\bvolume{11}(\bissue{1}),
\bfpage{1}--\blpage{11}
(\byear{2014})
\doiurl{10.1186/1743-0003-11-80/TABLES/3}
\end{barticle}
\endbibitem

\bibitem[\protect\citeauthoryear{Pratt et~al.}{2004}]{Pratt_2004}
\begin{barticle}
\bauthor{\bsnm{Pratt}, \binits{J.E.}},
\bauthor{\bsnm{Krupp}, \binits{B.T.}},
\bauthor{\bsnm{Morse}, \binits{C.J.}},
\bauthor{\bsnm{Collins}, \binits{S.H.}}:
\batitle{The roboknee: An exoskeleton for enhancing strength and endurance
  during walking}.
\bjtitle{Proceedings - IEEE International Conference on Robotics and
  Automation}
\bvolume{2004}(\bissue{3}),
\bfpage{2430}--\blpage{2435}
(\byear{2004})
\doiurl{10.1109/ROBOT.2004.1307425}
\end{barticle}
\endbibitem

\bibitem[\protect\citeauthoryear{Weinberg et~al.}{2007}]{Weinberg_2007}
\begin{botherref}
\oauthor{\bsnm{Weinberg}, \binits{B.}},
\oauthor{\bsnm{Nikitczuk}, \binits{J.}},
\oauthor{\bsnm{Patel}, \binits{S.}},
\oauthor{\bsnm{Patritti}, \binits{B.}},
\oauthor{\bsnm{Mavroidis}, \binits{C.}},
\oauthor{\bsnm{Bonato}, \binits{P.}},
\oauthor{\bsnm{Canavan}, \binits{P.}}:
Design, control and human testing of an active knee rehabilitation orthotic
  device.
Proceedings - IEEE International Conference on Robotics and Automation,
4126--4133
(2007)
\doiurl{10.1109/ROBOT.2007.364113}
\end{botherref}
\endbibitem

\bibitem[\protect\citeauthoryear{Bacek et~al.}{2017}]{Bacek_2017}
\begin{botherref}
\oauthor{\bsnm{Bacek}, \binits{T.}},
\oauthor{\bsnm{Moltedo}, \binits{M.}},
\oauthor{\bsnm{Langlois}, \binits{K.}},
\oauthor{\bsnm{Prieto}, \binits{G.A.}},
\oauthor{\bsnm{Sanchez-Villamañan}, \binits{M.C.}},
\oauthor{\bsnm{Gonzalez-Vargas}, \binits{J.}},
\oauthor{\bsnm{Vanderborght}, \binits{B.}},
\oauthor{\bsnm{Lefeber}, \binits{D.}},
\oauthor{\bsnm{Moreno}, \binits{J.C.}}:
Biomot exoskeleton - towards a smart wearable robot for symbiotic human-robot
  interaction.
IEEE International Conference on Rehabilitation Robotics,
1666--1671
(2017)
\doiurl{10.1109/ICORR.2017.8009487}
\end{botherref}
\endbibitem

\bibitem[\protect\citeauthoryear{Beyl et~al.}{2009}]{Beyl_2009}
\begin{botherref}
\oauthor{\bsnm{Beyl}, \binits{P.}},
\oauthor{\bsnm{Damme}, \binits{M.}},
\oauthor{\bsnm{Ham}, \binits{R.}},
\oauthor{\bsnm{Vanderborght}, \binits{B.}},
\oauthor{\bsnm{Lefeber}, \binits{D.}}:
Design and control of a lower limb exoskeleton for robot-assisted gait
  training.
Applied Bionics and Biomechanics
(2),
229--243
(2009)
\doiurl{10.1080/11762320902784393}
\end{botherref}
\endbibitem

\bibitem[\protect\citeauthoryear{Jezernik et~al.}{2003}]{Jezernik_2003}
\begin{barticle}
\bauthor{\bsnm{Jezernik}, \binits{S.}},
\bauthor{\bsnm{Colombo}, \binits{G.}},
\bauthor{\bsnm{Keller}, \binits{T.}},
\bauthor{\bsnm{Frueh}, \binits{H.}},
\bauthor{\bsnm{Morari}, \binits{M.}}:
\batitle{Robotic orthosis lokomat: A rehabilitation and research tool}.
\bjtitle{Neuromodulation: Technology at the Neural Interface}
\bvolume{6}(\bissue{2}),
\bfpage{108}--\blpage{115}
(\byear{2003})
\doiurl{10.1046/J.1525-1403.2003.03017.X}
\end{barticle}
\endbibitem

\bibitem[\protect\citeauthoryear{Van Der~Kooij et~al.}{2006}]{Van_2006}
\begin{botherref}
\oauthor{\bsnm{Van Der~Kooij}, \binits{H.}},
\oauthor{\bsnm{Veneman}, \binits{J.}},
\oauthor{\bsnm{Ekkelenkamp}, \binits{R.}}:
Design of a compliantly actuated exo-skeleton for an impedance controlled gait
  trainer robot.
Annual International Conference of the IEEE Engineering in Medicine and Biology
  - Proceedings,
189--193
(2006)
\doiurl{10.1109/IEMBS.2006.259397}
\end{botherref}
\endbibitem

\bibitem[\protect\citeauthoryear{}{2022}]{Mirad}
\begin{botherref}
Mirad.
\url{https://www.mirad-sbo.be/}.
Accessed: Mar. 14, 2022
(2022)
\end{botherref}
\endbibitem

\bibitem[\protect\citeauthoryear{}{2022}]{RexBionics}
\begin{botherref}
RexBionics.
\url{https://www.rexbionics.com/}.
Accessed: Mar. 14, 2022
(2022)
\end{botherref}
\endbibitem

\bibitem[\protect\citeauthoryear{Schiele}{2008}]{Schiele_2008}
\begin{botherref}
\oauthor{\bsnm{Schiele}, \binits{A.}}:
An explicit model to predict and interpret constraint force creation in phri
  with exoskeletons.
Proceedings - IEEE International Conference on Robotics and Automation,
1324--1330
(2008)
\doiurl{10.1109/ROBOT.2008.4543387}
\end{botherref}
\endbibitem

\bibitem[\protect\citeauthoryear{Schiele}{2009}]{Schiele_2009}
\begin{botherref}
\oauthor{\bsnm{Schiele}, \binits{A.}}:
Ergonomics of exoskeletons: Objective performance metrics.
Proceedings - 3rd Joint EuroHaptics Conference and Symposium on Haptic
  Interfaces for Virtual Environment and Teleoperator Systems, World Haptics
  2009,
103--108
(2009)
\doiurl{10.1109/WHC.2009.4810871}
\end{botherref}
\endbibitem

\bibitem[\protect\citeauthoryear{Morris et~al.}{2011}]{Morris_2011}
\begin{barticle}
\bauthor{\bsnm{Morris}, \binits{E.A.}},
\bauthor{\bsnm{Shorter}, \binits{K.A.}},
\bauthor{\bsnm{Li}, \binits{Y.}},
\bauthor{\bsnm{Hsiao-Wecksler}, \binits{E.T.}},
\bauthor{\bsnm{Kogler}, \binits{G.F.}},
\bauthor{\bsnm{Bretl}, \binits{T.}},
\bauthor{\bsnm{Durfee}, \binits{W.K.}}:
\batitle{Actuation timing strategies for a portable powered ankle foot
  orthosis}.
\bjtitle{ASME 2011 Dynamic Systems and Control Conference and Bath/ASME
  Symposium on Fluid Power and Motion Control, DSCC 2011}
\bvolume{2},
\bfpage{807}--\blpage{814}
(\byear{2011})
\doiurl{10.1115/DSCC2011-6170}
\end{barticle}
\endbibitem

\bibitem[\protect\citeauthoryear{Rastegarpanah
  et~al.}{2016}]{Rastegarpanah_2016}
\begin{barticle}
\bauthor{\bsnm{Rastegarpanah}, \binits{A.}},
\bauthor{\bsnm{Saadat}, \binits{M.}},
\bauthor{\bsnm{Borboni}, \binits{A.}}:
\batitle{Parallel robot for lower limb rehabilitation exercises}.
\bjtitle{Applied Bionics and Biomechanics}
(\byear{2016})
\doiurl{10.1155/2016/8584735}
\end{barticle}
\endbibitem

\bibitem[\protect\citeauthoryear{Chablat and Wenger}{1998}]{Chablat_1998}
\begin{barticle}
\bauthor{\bsnm{Chablat}, \binits{D.}},
\bauthor{\bsnm{Wenger}, \binits{P.}}:
\batitle{Working modes and aspects in fully parallel manipulators}.
\bjtitle{Proceedings - IEEE International Conference on Robotics and
  Automation}
\bvolume{3},
\bfpage{1964}--\blpage{1969}
(\byear{1998})
\doiurl{10.1109/ROBOT.1998.680601}
\end{barticle}
\endbibitem

\bibitem[\protect\citeauthoryear{Saglia et~al.}{2019}]{Saglia_2019}
\begin{botherref}
\oauthor{\bsnm{Saglia}, \binits{J.A.}},
\oauthor{\bsnm{Luca}, \binits{A.D.}},
\oauthor{\bsnm{Squeri}, \binits{V.}},
\oauthor{\bsnm{Ciaccia}, \binits{L.}},
\oauthor{\bsnm{Sanfilippo}, \binits{C.}},
\oauthor{\bsnm{Ungaro}, \binits{S.}},
\oauthor{\bsnm{Michieli}, \binits{L.D.}}:
Design and development of a novel core, balance and lower limb rehabilitation
  robot: Hunova®.
IEEE International Conference on Rehabilitation Robotics,
417--422
(2019)
\doiurl{10.1109/ICORR.2019.8779531}
\end{botherref}
\endbibitem

\bibitem[\protect\citeauthoryear{Girone et~al.}{2001}]{Girone_2001}
\begin{barticle}
\bauthor{\bsnm{Girone}, \binits{M.}},
\bauthor{\bsnm{Burdea}, \binits{G.}},
\bauthor{\bsnm{Bouzit}, \binits{M.}},
\bauthor{\bsnm{Popescu}, \binits{V.}},
\bauthor{\bsnm{Deutsch}, \binits{J.E.}}:
\batitle{A stewart platform-based system for ankle telerehabilitation}.
\bjtitle{Autonomous Robots 2001 10:2}
\bvolume{10}(\bissue{2}),
\bfpage{203}--\blpage{212}
(\byear{2001})
\doiurl{10.1023/A:1008938121020}
\end{barticle}
\endbibitem

\bibitem[\protect\citeauthoryear{Deuschl et~al.}{2020}]{Deuschl_2020}
\begin{botherref}
\oauthor{\bsnm{Deuschl}, \binits{G.}},
\oauthor{\bsnm{Beghi}, \binits{E.}},
\oauthor{\bsnm{Fazekas}, \binits{F.}},
\oauthor{\bsnm{Varga}, \binits{T.}},
\oauthor{\bsnm{Christoforidi}, \binits{K.A.}},
\oauthor{\bsnm{Sipido}, \binits{E.}},
\oauthor{\bsnm{Bassetti}, \binits{C.L.}},
\oauthor{\bsnm{Vos}, \binits{T.}},
\oauthor{\bsnm{Feigin}, \binits{V.L.}}:
The burden of neurological diseases in europe: an analysis for the global
  burden of disease study 2017.
The Lancet Public Health
\textbf{5}
(2020)
\doiurl{10.1016/S2468-2667(20)30190-0}
\end{botherref}
\endbibitem

\bibitem[\protect\citeauthoryear{Deutsch et~al.}{2001}]{Deutsch_2001b}
\begin{barticle}
\bauthor{\bsnm{Deutsch}, \binits{J.E.}},
\bauthor{\bsnm{Latonio}, \binits{J.}},
\bauthor{\bsnm{Burdea}, \binits{G.C.}},
\bauthor{\bsnm{Boian}, \binits{R.}}:
\batitle{Post-stroke rehabilitation with the rutgers ankle system: A case
  study}.
\bjtitle{Presence: Teleoperators and Virtual Environments}
\bvolume{10},
\bfpage{416}--\blpage{430}
(\byear{2001})
\doiurl{10.1162/1054746011470262}
\end{barticle}
\endbibitem

\bibitem[\protect\citeauthoryear{Cioi et~al.}{2011}]{Cioi_2011}
\begin{barticle}
\bauthor{\bsnm{Cioi}, \binits{D.}},
\bauthor{\bsnm{Kale}, \binits{A.}},
\bauthor{\bsnm{Burdea}, \binits{G.}},
\bauthor{\bsnm{Engsberg}, \binits{J.}},
\bauthor{\bsnm{Janes}, \binits{W.}},
\bauthor{\bsnm{Ross}, \binits{S.}}:
\batitle{Ankle control and strength training for children with cerebral palsy
  using the rutgers}.
\bjtitle{IEEE International Conference on Rehabilitation Robotics}
(\byear{2011})
\doiurl{10.1109/ICORR.2011.5975432}
\end{barticle}
\endbibitem

\bibitem[\protect\citeauthoryear{Girone et~al.}{2000}]{Girone_2000}
\begin{barticle}
\bauthor{\bsnm{Girone}, \binits{M.}},
\bauthor{\bsnm{Burdea}, \binits{G.}},
\bauthor{\bsnm{Bouzit}, \binits{M.}},
\bauthor{\bsnm{Popescu}, \binits{V.}},
\bauthor{\bsnm{Deutsch}, \binits{J.E.}}:
\batitle{Orthopedic rehabilitation using the “rutgers ankle” interface}.
\bjtitle{Studies in Health Technology and Informatics}
\bvolume{70},
\bfpage{89}--\blpage{95}
(\byear{2000})
\doiurl{10.3233/978-1-60750-914-1-89}
\end{barticle}
\endbibitem

\bibitem[\protect\citeauthoryear{}{2022}]{OptiFlex}
\begin{botherref}
OptiFlex.
\url{https://www.tru-medical.com/Modalities/OptiFlex-Ankle-CPM}.
Accessed: Jul. 03, 2022
(2022)
\end{botherref}
\endbibitem

\bibitem[\protect\citeauthoryear{}{2022}]{Breva}
\begin{botherref}
Breva.
\url{https://www.kinetecusa.com/shop/cpm-active/ankle-cpms/kinetec-breva/}.
Accessed: Jul. 03, 2022
(2022)
\end{botherref}
\endbibitem

\bibitem[\protect\citeauthoryear{Prosperini and
  Pozzilli}{2013}]{Prosperini_2013}
\begin{barticle}
\bauthor{\bsnm{Prosperini}, \binits{L.}},
\bauthor{\bsnm{Pozzilli}, \binits{C.}}:
\batitle{The clinical relevance of force platform measures in multiple
  sclerosis: A review}.
\bjtitle{Multiple Sclerosis International}
\bvolume{2013},
\bfpage{1}--\blpage{9}
(\byear{2013})
\doiurl{10.1155/2013/756564}
\end{barticle}
\endbibitem

\bibitem[\protect\citeauthoryear{Park and Lee}{2014}]{Park_2014}
\begin{botherref}
\oauthor{\bsnm{Park}, \binits{D.S.}},
\oauthor{\bsnm{Lee}, \binits{G.}}:
Validity and reliability of balance assessment software using the nintendo wii
  balance board: Usability and validation.
Journal of NeuroEngineering and Rehabilitation
\textbf{11}
(2014)
\doiurl{10.1186/1743-0003-11-99}
\end{botherref}
\endbibitem

\bibitem[\protect\citeauthoryear{}{2021}]{Wii}
\begin{botherref}
Wii Fit Plus.
\url{https://www.nintendo.co.uk/Games/Wii/Wii-Fit-Plus-283905.html}.
Accessed: Mar. 18, 2021
(2021)
\end{botherref}
\endbibitem

\bibitem[\protect\citeauthoryear{}{2022}]{Bertec}
\begin{botherref}
Force Plates.
\url{https://www.bertec.com/products/force-plates}.
Accessed: Jul. 03, 2022
(2022)
\end{botherref}
\endbibitem

\bibitem[\protect\citeauthoryear{}{2022}]{Kistler}
\begin{botherref}
Force Plate.
\url{https://www.kistler.com/en/glossary/term/force-plate/}.
Accessed: Jul. 03, 2022
(2022)
\end{botherref}
\endbibitem

\bibitem[\protect\citeauthoryear{}{2022}]{HURLabs}
\begin{botherref}
HURLabs.
\url{http://www.hurlabs.com/force-platform-fp}.
Accessed: Jul. 03, 2022
(2022)
\end{botherref}
\endbibitem

\bibitem[\protect\citeauthoryear{}{2022}]{AMTI}
\begin{botherref}
AMTI.
\url{https://www.amti.biz/product-line/force-plates/}.
Accessed: Jul. 03, 2022
(2022)
\end{botherref}
\endbibitem

\bibitem[\protect\citeauthoryear{Plow and Finlayson}{2014}]{Plow_2014}
\begin{barticle}
\bauthor{\bsnm{Plow}, \binits{M.}},
\bauthor{\bsnm{Finlayson}, \binits{M.}}:
\batitle{A qualitative study exploring the usability of nintendo wii fit among
  persons with multiple sclerosis}.
\bjtitle{Occupational Therapy International}
\bvolume{21}(\bissue{1}),
\bfpage{21}--\blpage{32}
(\byear{2014})
\doiurl{10.1002/oti.1345}
\end{barticle}
\endbibitem

\bibitem[\protect\citeauthoryear{}{2021}]{Balance}
\begin{botherref}
Balance Board Benefit.
\url{https://www.livestrong.com/article/34421-balance-board-benefits/}.
Accessed: Sep. 07, 2021
(2021)
\end{botherref}
\endbibitem

\bibitem[\protect\citeauthoryear{}{2021}]{bobo-balance}
\begin{botherref}
bobo balance.
\url{https://bobo-balance.com/}.
Accessed: Aug. 17, 2021
(2021)
\end{botherref}
\endbibitem

\bibitem[\protect\citeauthoryear{}{2021}]{Vertigomedprodotti}
\begin{botherref}
GeaPro.
\url{http://vertigomed.it/uk/index.php/vertigomed-prodotti/geapro/}.
Accessed: Aug. 17, 2021
(2021)
\end{botherref}
\endbibitem

\bibitem[\protect\citeauthoryear{}{2022}]{RiabloEuleria}
\begin{botherref}
Euleria.
\url{https://euleria.it/en/riablo-clinic/}.
Accessed: Feb. 03, 2022
(2022)
\end{botherref}
\endbibitem

\bibitem[\protect\citeauthoryear{}{2022}]{Balanceback}
\begin{botherref}
Balanceback.
\url{https://www.balanceback.com/}.
Accessed: Feb. 03, 2022
(2022)
\end{botherref}
\endbibitem

\bibitem[\protect\citeauthoryear{}{2022}]{Proprio}
\begin{botherref}
Proprio.
\url{https://www.perrydynamics.com/}.
Accessed: Feb. 23, 2022
(2022)
\end{botherref}
\endbibitem

\bibitem[\protect\citeauthoryear{}{2022}]{HUBER}
\begin{botherref}
HUBER.
\url{https://www.lpgmedical.com/en/professional-area/huber/}.
Accessed: Jul. 03, 2022
(2022)
\end{botherref}
\endbibitem

\bibitem[\protect\citeauthoryear{Kang et~al.}{2015}]{Kang_2015}
\begin{botherref}
\oauthor{\bsnm{Kang}, \binits{J.}},
\oauthor{\bsnm{Vashista}, \binits{V.}},
\oauthor{\bsnm{Agrawal}, \binits{S.K.}}:
A novel assist-as-needed control method to guide pelvic trajectory for gait
  rehabilitation.
IEEE International Conference on Rehabilitation Robotics,
630--635
(2015)
\doiurl{10.1109/ICORR.2015.7281271}
\end{botherref}
\endbibitem

\bibitem[\protect\citeauthoryear{Negahban et~al.}{2011}]{Negahban_2011}
\begin{barticle}
\bauthor{\bsnm{Negahban}, \binits{H.}},
\bauthor{\bsnm{Mofateh}, \binits{R.}},
\bauthor{\bsnm{Arastoo}, \binits{A.A.}},
\bauthor{\bsnm{Mazaheri}, \binits{M.}},
\bauthor{\bsnm{Yazdi}, \binits{M.J.S.}},
\bauthor{\bsnm{Salavati}, \binits{M.}},
\bauthor{\bsnm{Majdinasab}, \binits{N.}}:
\batitle{The effects of cognitive loading on balance control in patients with
  multiple sclerosis}.
\bjtitle{Gait and Posture}
\bvolume{34}(\bissue{4}),
\bfpage{479}--\blpage{484}
(\byear{2011})
\doiurl{10.1016/j.gaitpost.2011.06.023}
\end{barticle}
\endbibitem

\bibitem[\protect\citeauthoryear{Bottaro}{2008}]{Bottaro_2008}
\begin{barticle}
\bauthor{\bsnm{Bottaro}, \binits{B.} \bsuffix{Larsen}}:
\batitle{Neural correlates supporting sensory discrimination after left
  hemisphere stroke}.
\bjtitle{Bone}
\bvolume{23}(\bissue{1}),
\bfpage{1}--\blpage{7}
(\byear{2008})
\doiurl{10.1016/j.brainres.2012.03.060.Neural}
\end{barticle}
\endbibitem

\bibitem[\protect\citeauthoryear{Rossignol et~al.}{2006}]{Rossignol_2006}
\begin{botherref}
\oauthor{\bsnm{Rossignol}, \binits{S.}},
\oauthor{\bsnm{Dubuc}, \binits{R.}},
\oauthor{\bsnm{Gossard}, \binits{J.P.}}:
Dynamic sensorimotor interactions in locomotion.
Physiological Reviews,
89--154
(2006)
\doiurl{10.1152/physrev.00028.2005}
\end{botherref}
\endbibitem

\bibitem[\protect\citeauthoryear{Laaksonen et~al.}{2012}]{Laaksonen_2012}
\begin{barticle}
\bauthor{\bsnm{Laaksonen}, \binits{K.}},
\bauthor{\bsnm{Kirveskari}, \binits{E.}},
\bauthor{\bsnm{Mäkelä}, \binits{J.P.}},
\bauthor{\bsnm{Kaste}, \binits{M.}},
\bauthor{\bsnm{Mustanoja}, \binits{S.}},
\bauthor{\bsnm{Nummenmaa}, \binits{L.}},
\bauthor{\bsnm{Tatlisumak}, \binits{T.}},
\bauthor{\bsnm{Forss}, \binits{N.}}:
\batitle{Effect of afferent input on motor cortex excitability during stroke
  recovery}.
\bjtitle{Clinical Neurophysiology}
\bvolume{123}(\bissue{12}),
\bfpage{2429}--\blpage{2436}
(\byear{2012})
\doiurl{10.1016/j.clinph.2012.05.017}
\end{barticle}
\endbibitem

\bibitem[\protect\citeauthoryear{Lee et~al.}{2018}]{Lee_2018}
\begin{barticle}
\bauthor{\bsnm{Lee}, \binits{B.C.}},
\bauthor{\bsnm{Fung}, \binits{A.}},
\bauthor{\bsnm{Thrasher}, \binits{T.A.}}:
\batitle{The effects of coding schemes on vibrotactile biofeedback for dynamic
  balance training in parkinson’s disease and healthy elderly individuals}.
\bjtitle{IEEE Transactions on Neural Systems and Rehabilitation Engineering}
\bvolume{26},
\bfpage{153}--\blpage{160}
(\byear{2018})
\doiurl{10.1109/TNSRE.2017.2762239}
\end{barticle}
\endbibitem

\bibitem[\protect\citeauthoryear{Hasan and Dhingra}{2020}]{Hasan_2020}
\begin{barticle}
\bauthor{\bsnm{Hasan}, \binits{S.K.}},
\bauthor{\bsnm{Dhingra}, \binits{A.K.}}:
\batitle{State of the art technologies for exoskeleton human lower extremity
  rehabilitation robots}.
\bjtitle{Journal of Mechatronics and Robotics}
\bvolume{4}(\bissue{1}),
\bfpage{211}--\blpage{235}
(\byear{2020})
\doiurl{10.3844/jmrsp.2020.211.235}
\end{barticle}
\endbibitem

\bibitem[\protect\citeauthoryear{Sport and
  Education}{2002}]{Sport_Education_2002}
\begin{bbook}
\bauthor{\bsnm{Sport}, \binits{S.I.N.}},
\bauthor{\bsnm{Education}, \binits{P.}}:
\bbtitle{Jarmo Perttunen Foot Loading in Normal and Pathological Walking Foot
  Loading in Normal and Pathological Walking},
(\byear{2002})
\end{bbook}
\endbibitem

\bibitem[\protect\citeauthoryear{Jagdhane et~al.}{2016}]{Jagdhane_2016}
\begin{barticle}
\bauthor{\bsnm{Jagdhane}, \binits{S.}},
\bauthor{\bsnm{Kanekar}, \binits{N.}},
\bauthor{\bsnm{S.~Aruin}, \binits{A.}}:
\batitle{The effect of a four-week balance training program on anticipatory
  postural adjustments in older adults: A pilot feasibility study}.
\bjtitle{Current aging science}
\bvolume{9}(\bissue{4}),
\bfpage{295}--\blpage{300}
(\byear{2016})
\doiurl{10.2174/1874609809666160413113443}
\end{barticle}
\endbibitem

\bibitem[\protect\citeauthoryear{Aloraini et~al.}{2019}]{Aloraini_2019}
\begin{barticle}
\bauthor{\bsnm{Aloraini}, \binits{S.M.}},
\bauthor{\bsnm{Glazebrook}, \binits{C.M.}},
\bauthor{\bsnm{Sibley}, \binits{K.M.}},
\bauthor{\bsnm{Singer}, \binits{J.}},
\bauthor{\bsnm{Passmore}, \binits{S.}}:
\batitle{Anticipatory postural adjustments during a fitts’ task: Comparing
  young versus older adults and the effects of different foci of attention}.
\bjtitle{Human Movement Science}
\bvolume{64},
\bfpage{366}--\blpage{377}
(\byear{2019})
\doiurl{10.1016/j.humov.2019.02.019}
\end{barticle}
\endbibitem

\bibitem[\protect\citeauthoryear{Riemenschneider
  et~al.}{2018}]{Riemenschneider_2018}
\begin{barticle}
\bauthor{\bsnm{Riemenschneider}, \binits{M.}},
\bauthor{\bsnm{Hvid}, \binits{L.G.}},
\bauthor{\bsnm{Stenager}, \binits{E.}},
\bauthor{\bsnm{Dalgas}, \binits{U.}}:
\batitle{Is there an overlooked “window of opportunity” in ms exercise
  therapy? perspectives for early ms rehabilitation}.
\bjtitle{Multiple Sclerosis Journal}
\bvolume{24}(\bissue{7}),
\bfpage{886}--\blpage{894}
(\byear{2018})
\doiurl{10.1177/1352458518777377}
\end{barticle}
\endbibitem

\bibitem[\protect\citeauthoryear{Ilett et~al.}{2016}]{Ilett_2016}
\begin{barticle}
\bauthor{\bsnm{Ilett}, \binits{P.}},
\bauthor{\bsnm{Lythgo}, \binits{N.}},
\bauthor{\bsnm{Martin}, \binits{C.}},
\bauthor{\bsnm{Brock}, \binits{K.}}:
\batitle{Balance and gait in people with multiple sclerosis: A comparison with
  healthy controls and the immediate change after an intervention based on the
  bobath concept}.
\bjtitle{Physiotherapy Research International}
\bvolume{21}(\bissue{2}),
\bfpage{91}--\blpage{101}
(\byear{2016})
\doiurl{10.1002/pri.1624}
\end{barticle}
\endbibitem

\bibitem[\protect\citeauthoryear{Freyler et~al.}{2015}]{Freyler_2015}
\begin{botherref}
\oauthor{\bsnm{Freyler}, \binits{K.}},
\oauthor{\bsnm{Gollhofer}, \binits{A.}},
\oauthor{\bsnm{Colin}, \binits{R.}},
\oauthor{\bsnm{Brüderlin}, \binits{U.}},
\oauthor{\bsnm{Ritzmann}, \binits{R.}}:
Reactive balance control in response to perturbation in unilateral stance
  interaction effects of direction, displacement and velocity on compensatory
  neuromuscular and kinematic responses
\textbf{10}(2)
(2015)
\doiurl{10.1371/journal.pone.0144529}
\end{botherref}
\endbibitem

\bibitem[\protect\citeauthoryear{Krebs~DE.}{2001}]{krebs_2001}
\begin{botherref}
\oauthor{\bsnm{Krebs~DE.}, \binits{G.D.} \bsuffix{McGibbon~CA.}}:
Analysis of postural perturbation responses.
transactions on neural systems and rehabilitation engineering,
76--80
(2001)
\doiurl{10.1109/7333.918279}
\end{botherref}
\endbibitem

\end{thebibliography}

\end{document}